\documentclass[sigconf]{acmart}
\usepackage{multirow}
\usepackage{makecell}
\usepackage{subfig}
\usepackage{bbding}

\copyrightyear{2023}
\acmYear{2023}
\setcopyright{acmcopyright}
\acmConference[WSDM '23] {Proceedings of the Sixteenth ACM International Conference on Web Search and Data Mining}{February 27--March 3, 2023}{Singapore, Singapore.}
\acmBooktitle{Proceedings of the Sixteenth ACM International Conference on Web Search and Data Mining (WSDM '23), February 27--March 3, 2023, Singapore, Singapore}
\acmPrice{15.00}
\acmISBN{978-1-4503-9407-9/23/02}
\acmDOI{10.1145/3539597.3570372}
\begin{document}
\title{CL4CTR: A Contrastive Learning Framework for CTR Prediction}

\author{Fangye Wang}
\authornote{
Also Shanghai Key Laboratory of Data Science, Fudan University, Shanghai, China.
}
\orcid{0000-0001-7216-1688}
\affiliation{
{
\normalsize
  \institution{School of Computer Science}
  \institution{Fudan University
  \city{Shanghai}
  \country{China}}
}
}
\email{fywang18@fudan.edu.cn}

\author{Yingxu Wang}
\orcid{0000-0003-1208-9324}
\authornotemark[1]
\affiliation{
{
\normalsize
  \institution{School of Computer Science}
  \institution{Fudan University
  \city{Shanghai}
  \country{China}}
}
}
\email{yingxuwang20@fudan.edu.cn}

\author{Dongsheng Li}
\orcid{0000-0003-3103-8442}
\affiliation{
{
\normalsize
  \institution{Microsoft Research Asia}
  \city{Shanghai}
  \country{China}
 }
}
\email{dongsli@microsoft.com}

\author{Hansu Gu}
\authornote{Corresponding author.}
\orcid{0000-0002-1426-3210}
\affiliation{
{
\normalsize
  \city{Seattle}
  \country{United States}
}
}
\email{hansug@acm.org}

\author{Tun Lu}
\authornotemark[1]
\authornotemark[2]
% \authornote{Corresponding author.}
\orcid{0000-0002-6633-4826}
\affiliation{
\normalsize
  \institution{School of Computer Science}
  \institution{Fudan University 
  \city{Shanghai} 
  \country{China}}
 }
\email{lutun@fudan.edu.cn}

\author{Peng Zhang}
\authornotemark[1]
\orcid{0000-0002-9109-4625}
\affiliation{
 \normalsize
  \institution{School of Computer Science}
  \institution{Fudan University
  \city{Shanghai}
  \country{China}
  }
 }
\email{zhangpeng_@fudan.edu.cn}

\author{Ning Gu}
\authornotemark[1]
\orcid{0000-0002-2915-974X}
\affiliation{
{
  \normalsize
  \institution{School of Computer Science}
  \institution{Fudan University
  \city{Shanghai}
  \country{China}}
 }
}
\email{ninggu@fudan.edu.cn}

\renewcommand{\shortauthors}{Fangye Wang et al.}

\begin{abstract}
Many Click-Through Rate (CTR) prediction works focused on designing advanced architectures to model complex feature interactions but neglected the importance of feature representation learning, e.g., adopting a plain embedding layer for each feature, which results in sub-optimal feature representations and thus inferior CTR prediction performance. For instance, low frequency features, which account for the majority of features in many CTR tasks, are less considered in standard supervised learning settings, leading to sub-optimal feature representations. In this paper, we introduce self-supervised learning to produce high-quality feature representations directly and propose a model-agnostic Contrastive Learning for CTR (CL4CTR) framework consisting of three self-supervised learning signals to regularize the feature representation learning: contrastive loss, feature alignment, and field uniformity. The contrastive module first constructs positive feature pairs by data augmentation and then minimizes the distance between the representations of each positive feature pair by the contrastive loss. The feature alignment constraint forces the representations of features from the same field to be close, and the field uniformity constraint forces the representations of features from different fields to be distant. Extensive experiments verify that CL4CTR achieves the best performance on four datasets and has excellent effectiveness and compatibility with various representative baselines.
\end{abstract}



% \ccsdesc[500]{Information systems~Recommender systems}

\maketitle

\section{Introduction}

CTR prediction ~\cite{zhang2021survey_deep, guo2017deepfm}, aiming to predict the probability of a given item being clicked, has been widely used in many applications, e.g., recommender systems~\cite{cheng2016wide} and computational advertising ~\cite{liu2019feature}. Recently, many methods~\cite{chen2021enhancing,wang2022enhancing} achieved huge success by modeling complex feature interactions (FI). Following recent works ~\cite{guo2022miss, wang2022enhancing,huang2019fibinet}, we categorize CTR prediction methods into two types: (1) traditional methods, such as logistic regression (LR)~\cite{richardson2007predicting} and FM-based models ~\cite{rendle2012factorization,he2017neural}, can only model low-order feature interactions; and (2) deep-learning based methods, such as xDeepFM ~\cite{lian2018xdeepfm} and DCN-V2 \cite{wang2020dcn}, can further enhance the accuracy of CTR prediction by capturing high-order FI. In addition, many novel architectures (e.g., self-attention~\cite{song2019autoint,li2020interpretable}, CIN~\cite{lian2018xdeepfm}, PIN~\cite{qu2018product}) have been proposed and widely deployed to capture sophisticated arbitrary-order FI. 

\begin{figure}[tb]
\setlength{\abovecaptionskip}{0.2cm}
\setlength{\belowcaptionskip}{-0.2cm}
\centering
\label{fig:fre_cum}
\subfloat[Frappe]{
\begin{minipage}[t]{0.5\linewidth}
\centering
\includegraphics[width=0.90\textwidth]{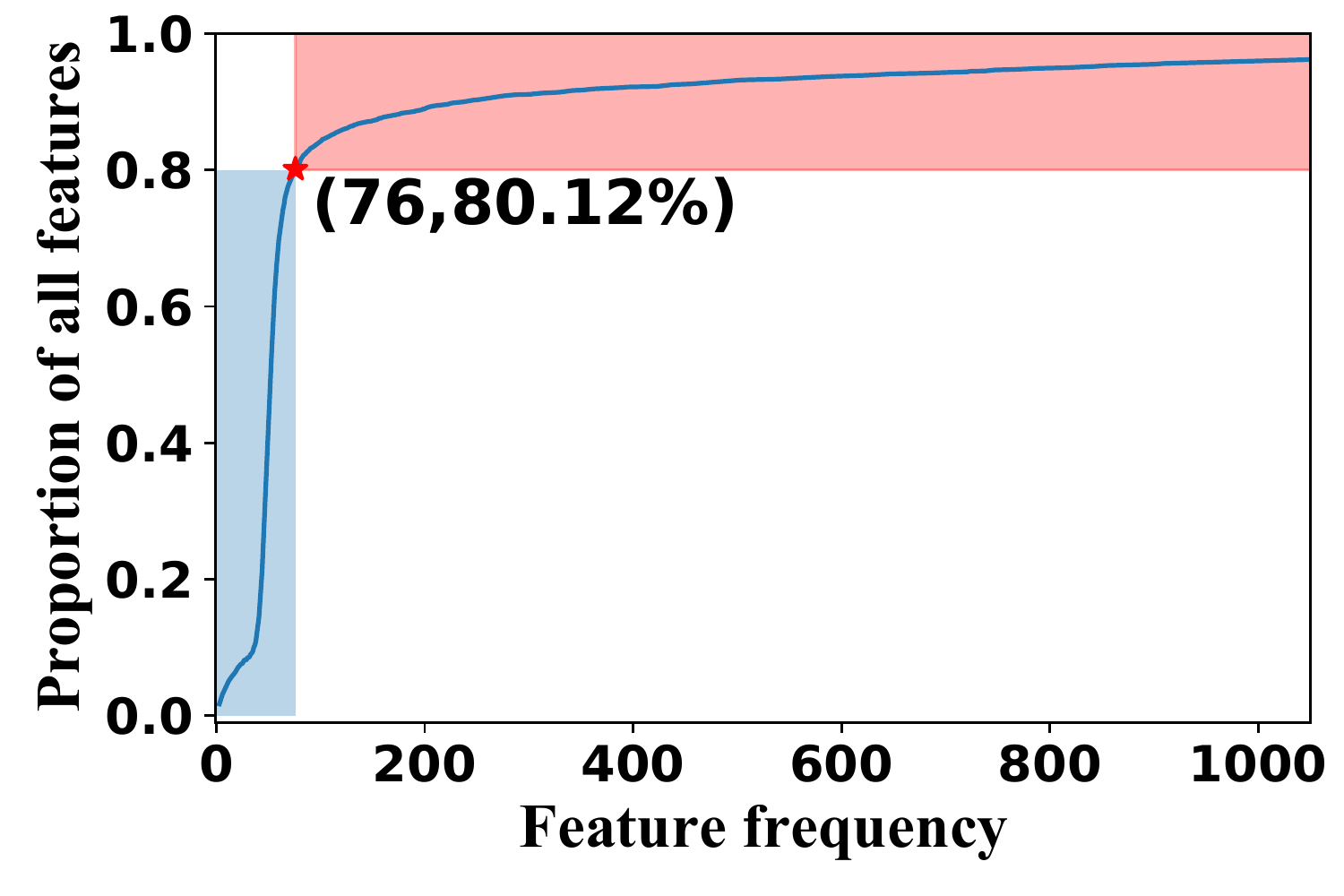}
\end{minipage}
}
\subfloat[ML-tag]{
\begin{minipage}[t]{0.5\linewidth}
\centering
\includegraphics[width=0.90\textwidth]{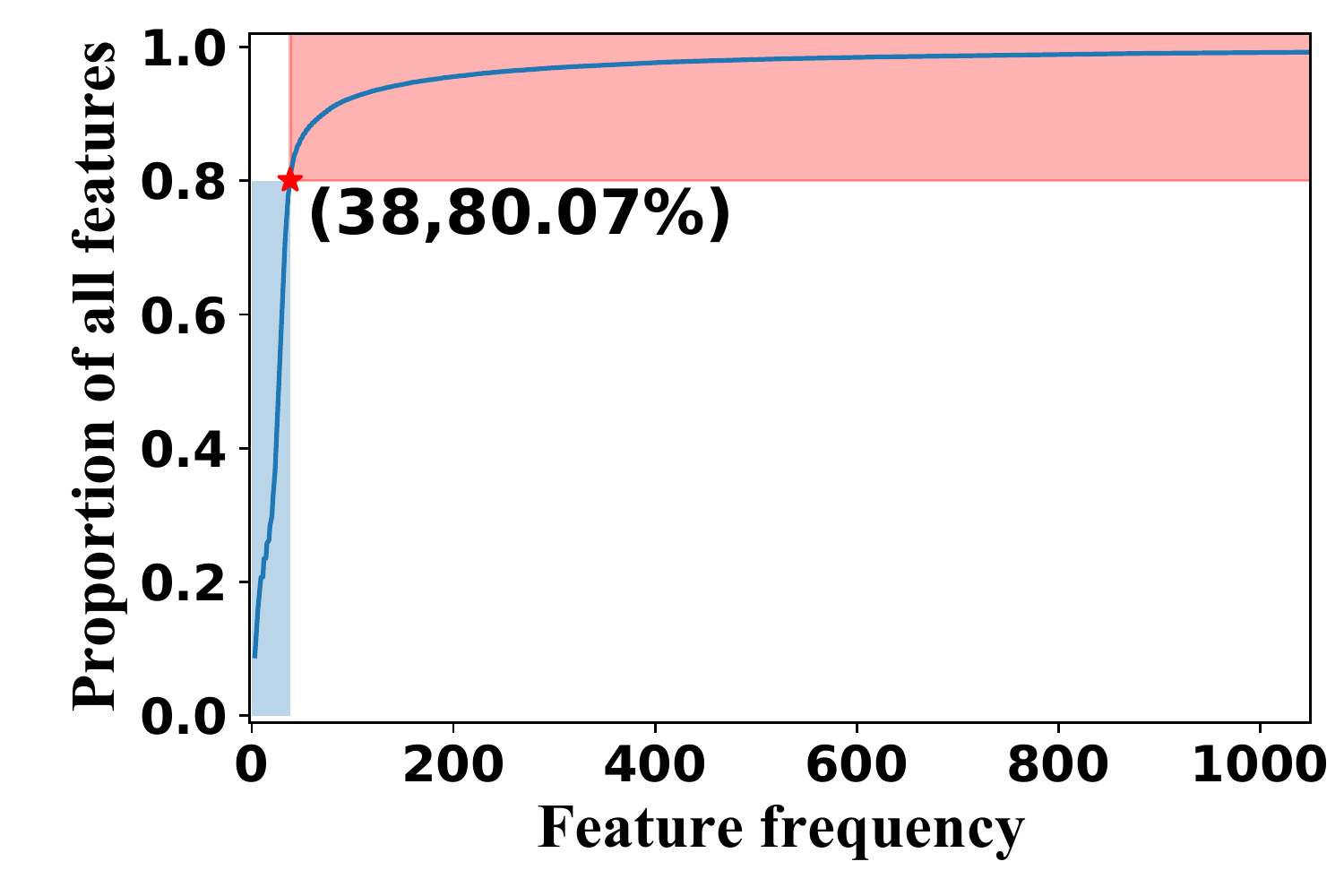}
\end{minipage}
}
\centering
\caption{Cumulative distribution of feature frequencies. (38, 80.07\%) indicates that 
features with feature frequencies less than or equal to 38 times account for 80.07\% of all features.}
\label{fig:fre_cum}
\end{figure}

Although successful in performance, many existing CTR prediction methods suffer from an inherent problem: {\em high frequency features have higher chances to be trained than low frequency features}, causing the representations of low frequency features to be sub-optimal. In Figure \ref{fig:fre_cum}, we present the feature cumulative distributions of Frappe and ML-tag datasets. We can observe a clear ``long tail'' distribution of feature frequencies, e.g., bottom 80\% of features appeared only 38 times or less in the ML-tag dataset. Since most CTR prediction models learn feature representations by the back-propagation~\cite{zhang2021survey_deep}, the low frequency features cannot be sufficiently trained due to less appearance, resulting in sub-optimal feature representations and thus sub-optimal CTR prediction performance. 

Several prior works~\cite{yu2019input,lu2020dual} have also realized the importance of feature representation learning and proposed to deploy a weight learning module (i.e., FEN~\cite{yu2019input}, Dual-FEN~\cite{lu2020dual}) after the embedding layer which assigns weights for each feature to enhance their representations. However, the additional weighting modules will increase the model parameters and inference time. In addition, similar to FI-based methods, these methods only use the supervised learning signals to optimize feature representations from the plain embedding layer, which is not strong enough to produce accurate feature representations. Therefore, in this paper, we focus on directly learning accurate feature representations from the embedding layer without introducing additional weighting mechanisms, which is model-agnostic and has not been extensively studied.

In this paper, we seek to utilize self-supervised learning (SSL) to address the above issue, in which we design self-supervised learning signals as constraints to regularize the learned feature representations during the training process. As shown in Figure~\ref{fig:model_cl4ctr}, we propose a novel framework called Contrastive Learning for Click Through Rate Prediction (\textbf{CL4CTR}), which consists of three key modules: CTR prediction model, contrastive module, and alignment\&uniformity constraints. In detail, the CTR prediction model aims to predict the probability of items being clicked by a user, which can be replaced with most existing CTR models in the CL4CTR framework. In the contrastive module, we design three key components: (1) a data augmentation unit aiming to generate two different views for the output embedding as positive training pairs, which includes three different permutation approaches: random mask, feature mask, and dimension mask; (2) a feature interaction encoder aiming to learn compact FI representations based on the perturbed embeddings from the data augmentation unit; and (3) a task-oriented contrastive loss, which is designed to minimize the distance between the positive training pairs. In addition, we introduce two constraints: feature alignment and field uniformity, to facilitate contrastive learning. Feature alignment forces the representations of features from the same field to be as close as possible, and field uniformity forces the representations of features from different fields to be as distant as possible. 

Our major contributions are summarized as follows:

\begin{itemize}
    \item We propose a model-agnostic contrastive learning framework -- CL4CTR, which can directly improve the quality of feature representations in an end-to-end manner. 
    \item Considering the unique characteristics of CTR prediction tasks, we design three self-supervised learning signals: contrastive loss, feature alignment constraint and field uniformity constraint to improve contrastive learning performance. 
    \item Extensive experiments on four datasets demonstrate that simply applying CL4CTR into FM~\cite{rendle2012factorization} can outperform state-of-the-art methods. More importantly, CL4CTR shows high compatibility with existing methods, i.e., it can generally improve the performance of many representative baselines.
\end{itemize}

\section{Related Work}
\subsection{Deep CTR Prediction}
According to the main focuses, recent CTR prediction works can be divided into two categories: FI-based methods~\cite{zhang2021survey_deep,wang2020dcn} and user interests modeling based methods~\cite{qin2020user}. Since our CL4CTR framework can be generally applied in FI-based models, we briefly summarize the FI-based works in this section. 
Most FI-based CTR prediction methods follow the common design paradigm: embedding layer, FI layer, and prediction layer. Some classical methods can only model fixed-order or low-order feature interactions. For instance, FM-based methods~\cite{rendle2012factorization,juan2016field,lu2020dual} model all pairwise interactions by using factorized parameters. Due to the importance of FI in the CTR prediction, many works focus on how to design novel structures for the FI layer to capture more informative and complicated feature interactions. Wide\&Deep (WDL)~\cite{cheng2016wide} jointly trains the wide linear unit and Deep Neural Network (DNN) to combine memorization and generalization. DeepFM~\cite{guo2017deepfm} comprises DNN and FM, and xDeepFM~\cite{lian2018xdeepfm} additional proposes Compressed Interaction Network (CIN) based DeepFM to model high-order feature interaction explicitly. DCN~\cite{wang2017deep} and DCN-V2~\cite{wang2020dcn} explicitly and automatically use a cross-vector/cross-matrix network to improve the accuracy and efficiency of the DNN model. Furthermore, attention mechanism is one of the most effective structure to improve the performance and has been widely adopted for different purposes, e.g., AFM~\cite{xiao2017attentional}, Autoint~\cite{song2019autoint}, InterCTR~\cite{li2020interpretable},  DCAP~\cite{chen2021dcap}.

Notably, some works~\cite{yu2019input,lu2020dual,huang2019fibinet} attempt to improve the performance of CTR prediction by assigning different weights for features, in which they deploy a weight learning module to adjust the importance of feature representations after the embedding layer. However, these additional weighting modules may increase the model parameters and inference time. More importantly, these works only learn feature representations from a plain embedding layer, which is not strong enough to produce accurate feature representations as demonstrated in our experiments. The proposed CL4CTR can directly improve the quality of feature representations without requiring any additional modules after the embedding layer.

\subsection{Self-supervised Learning}
Recently, self-supervised learning has achieved remarkable success in learning powerful representations in many machine learning tasks~\cite{chen2020simple, he2020momentum,gao2021simcse,xie2020CL4SREC,lee2021bootstrapping, verma2021towards}. Contrastive Learning  is one of the mainstream methods in SSL, which learns representations by attracting the positive sample pairs and repulsing the negative sample pairs~\cite{guo2022miss}. \citet{wang2020understanding} identify two key properties related to the success of contrastive learning, i.e., alignment and uniformity. Alignment favors encoders that assign similar features to similar samples. Uniformity prefers a feature distribution that preserves maximal information. 

In CTR prediction tasks, contrastive learning has not been extensively studied. ~\citet{guo2022miss} focus on sequential-based CTR tasks, which apply interest-level contrastive losses to enhance feature embeddings. ~\citet{pan2021aqclclick} propose an auxiliary AQCL loss that automatically leverages instance-instance similarity and instance-cluster similarity to regularize feature representations under the cold-start scenarios. Unlike them, our CL4CTR focuses on FI-based CTR prediction models, which can enhance the quality of feature representations by designing three SSL signals: contrastive loss, feature alignment constraint, and field uniformity constraint. 

\section{The CL4CTR Framework}
\label{model_cl4ctr}
\subsection{CTR Prediction}
\label{sec:pre}
CTR prediction is a binary classification task~\cite{wang2022enhancing, pan2021aqclclick}. Suppose a dataset for training CTR prediction model contains $N$ instances $(\mathbf{x},y)$, where $y\in{\{0,1\}}$ (click or not) is the true label indicating user’s click behaviors. Input instance $\mathbf{x}$ is usually multi-field tabular data record~\cite{guo2022miss, luo2020network}, which contains $F$ different fields and $M$ features, as shown in Table \ref{tab:tabular}. Recently, as shown in Figure \ref{fig:model_cl4ctr}(a), many CTR prediction models follow the common design paradigm~\cite{wang2021masknet, wang2022enhancing}: embedding layer, FI layer, and prediction layer.

\textbf{{Embedding layer.}} 
Generally, each input instance $\mathbf{x}_i$ is a sparse high-dimensional vector represented by a one-hot vector~\cite{wang2022enhancing, liu2019feature}. And embedding layer transforms the sparse high-dimensional features $\mathbf{x}_i$ into a dense low-dimensional embedding matrix $\mathbf{E}=[\mathbf{e}^1;\mathbf{e}^2;...;\mathbf{e}^F] \in \mathbb{R}^{F \times D}$, where $D$ is the dimension size. Additionally, we use $\mathbb{E} = [\mathbf{E}_{1},\mathbf{E}_{2},...,\mathbf{E}_{F}] \in \mathbb{R}^{M \times D}$ to represent all feature representations, where $\mathbf{E}_{f}$ is the subset representation of the $f$-th field $f \in \{1,2,...,F\}$. $\lvert \mathbf{E}_{f} \rvert$ is the number of features belonging to field $f$, and $M = \sum_{f=1}^{F} {\lvert \mathbf{E}_{f} \rvert}$.

\textbf{{Feature interaction layer.}} 
\label{sec:fi_cncoder}
The FI layer usually contains various types of interaction operations to capture arbitrary-order feature interactions, such as MLP~\cite{guo2017deepfm,cheng2016wide}, Cross Network~\cite{wang2017deep}, Cross Network2~\cite{wang2020dcn} and transformer layer~\cite{li2020interpretable,song2019autoint}, etc. We refer to these structures as feature interaction encoders, represented by $FI(\cdot)$. $FI(\cdot)$ can generate a compact feature interaction representation $\mathbf{\mathit{h}}_i$ based on embedding matrix $\mathbf{E}$. 

\textbf{{Prediction layer.}} Finally, a prediction layer (usually a linear regression or MLP module) produces the final prediction probability $\sigma(\hat{y_i}) \in [0, 1]$ based on the compact representations $\mathbf{\mathit{h}}_i$ from the FI layer, where $\sigma(x)=1 /(1+\exp (-x))$ is the sigmoid function. 

Finally, with the predicted label $\hat{y}_i$ and the true label $y_i$, the commonly adopted loss function of CTR models is as follows:
\begin{align}
\textstyle
\mathcal{L}_{ctr} =-\frac{1}{N}\sum_{i=1}^N 
\left({y_i} 
    \log\left(\sigma \left(\hat{y}_i \right)\right) 
    +\left( 1-y_i \right) \log \left( 1-\sigma \left(\hat{y}_i \right)\right)
\right).
\end{align}

\textbf{{Contrastive learning.}} As shown in Figure~\ref{fig:model_cl4ctr}, in addition to the above components, we propose three contrastive learning signals: contrastive loss, feature alignment constraint and field uniformity constraint on top of the embedding layer to regularize the representation learning. Since these signals are not necessary during model inference, our method will not increase the inference time and the parameters of the underlying CTR prediction models.

\begin{table}
\centering
\caption{An example of multi-field tabular data for CTR prediction. Each row represents an input instance and each column indicates a field. Moreover, each field contains multiple features, but each feature only belongs to one field.}
\label{tab:tabular}
\scalebox{0.90}{\begin{tabular}{ccccc|c|c} 
\toprule
user\_id & item\_id & gender & city & daytime & ... & click  \\ 
\hline
25c83c98      & c5c50484& female      & 5    & 1       & ... & 0\\
7e0ccccf      & 0b153874& male       & 10   & 5       & ... & 1      \\
de7995b8      & e51ddf94& male       & 32   & 6       & ... & 1      \\
1f89b562      & f0cf0024& female     & 4    & 2       & ... & 1      \\
1f89b562      & a3397841& female      & 4    & 8       & ... & 0      \\
\bottomrule
\end{tabular}
}
\end{table}

\subsection{Contrastive Module }
\begin{figure*}[t]
    \setlength{\abovecaptionskip}{0.2cm}
    \setlength{\belowcaptionskip}{-0.2cm}
    \centering
    \includegraphics[width=0.80\textwidth]{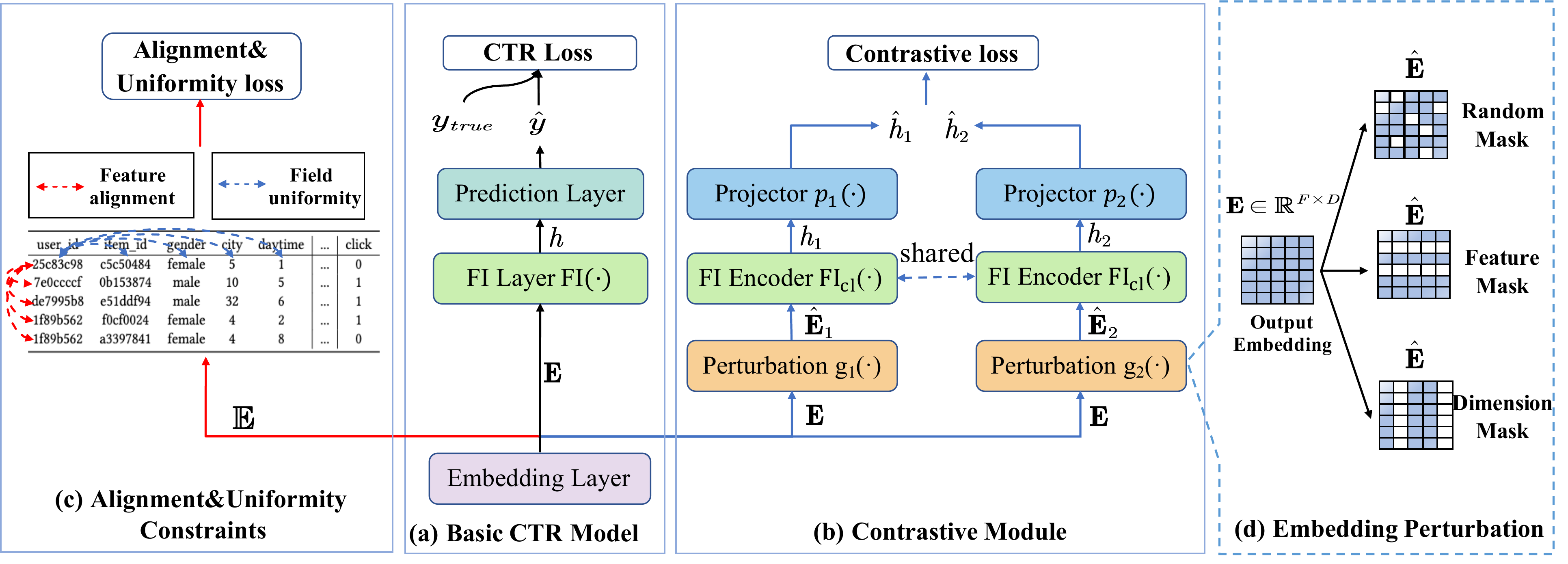}
    \caption{Architecture of the CL4CTR framework. CL4CTR including three components: (a) a basic CTR model; (b) a contrastive module; (c)  alignment \& uniformity constraints. In contrastive module, we design (d) three embedding perturbation methods.}
    \label{fig:model_cl4ctr}
\end{figure*}

Inspired by the success of SSL, we seek to deploy contrastive learning in the CTR prediction tasks to generate high-quality feature representations. As illustrated in Figure \ref{fig:model_cl4ctr}(b), the contrastive module consists of three major components: a data augmentation unit, a FI encoder, and a contrastive loss function. In the data augmentation unit, we propose three different task-oriented posterior embedding augmentation techniques to generate positive training pairs, i.e., two different views of each feature embedding. Then we feed the two perturbed embeddings to the same FI encoder to generate two compressed feature representations. Finally, the contrastive loss is applied to minimize the distance between the two compressed feature representations. 

\subsubsection{Data Augmentation via Output Perturbation}

Data augmentation has shown great potential in improving the performance of feature representations in SSL~\cite{yu2022self}. Different and well-designed augmentation approaches have been proposed and used to construct different views of the same input instance. For example, in the scenarios of sequential recommendation, three widely used augmentation methods are item masking, reordering, and cropping~\cite{yu2022self}. However, these methods are designed to augment behavior sequences and are not appropriately deployed in FI-based CTR prediction models. Hence, we firstly propose three task-oriented augmentation approaches, which aim to perturb feature embeddings for FI-based models. As shown in Figure 2(d), we use the function $\hat{\mathbf{E}} = \mathrm{g}(\mathbf{E})$ to represent data augmentation process.

\textbf{Random Mask.} Firstly, we introduce the random mask method, which is analogous to Dropout~\cite{hinton2012improving}. This method randomly masks some elements in initial embedding $\textbf{E}$ with a certain probability \textit{p}. The  random mask is generated as follows:
\begin{align}
    \hat{\mathbf{E}} = \mathrm{g}_r(\mathbf{E}) = \mathbf{E} \cdot \mathbf{I},
    \mathbf{I} \sim \operatorname{Bernoulli}(p) \in \mathbb{E}^{F \times D}.
\end{align}
$\operatorname{Bernoulli(\cdot)}$ is the Bernoulli distribution, and $\mathbf{I}$ is a matrix of Bernoulli random variables each of which has probability $p$ of being 1.

\textbf{Feature Mask.}
Motivated by prior works~\cite{liu2019feature,song2019autoint}, we propose to mask the feature information in the initial embedding,  where the feature mask can be generated as follows: 
\begin{align}
    \hat{\mathbf{E}} = \mathrm{g}_f(\mathbf{E}) = [\hat{\mathbf{e}}^{1};\hat{\mathbf{e}}^{2};...;\hat{\mathbf{e}}^{F}], 
    \hat{\mathbf{e}}^{f} = 
    \begin{cases} 
        \mathbf{e}^{f}, & t \notin \mathcal{T} \\ {[\mathrm{mask}],} & t \in \mathcal{T}
    \end{cases},
\end{align}
where we set a proportion $p$ of features $\mathcal{T} = (t_1, t_2,..., t_{L_F})$ with the length $L_F = \left\lfloor p *F\right\rfloor$. $t_f$ is the index of feature in $\mathbf{E}$. If one feature is masked, then the representation of this feature will be replaced with [mask], which is a zero vector.

\textbf{Dimension Mask.}
The dimensions of feature representations affect the effectiveness of deep learning models. Inspired by FED~\cite{zhao2020fed_dimension}, which attempts to improve prediction performance by capturing dimension relations, we propose to perturb the initial embedding by replacing specific proportions of dimensional information of feature representations, which can be described as follows:
\begin{align}
\textstyle
    \hat{\mathbf{E}} = \mathrm{g}_d(\mathbf{E}) = [d\mathbf{e}^{1};d\mathbf{e}^{2};...;d\mathbf{e}^{F}], 
    d \sim \operatorname{Bernoulli}(p) \in \mathbb{R}^{D},
\end{align}
where $d$ is a vector of Bernoulli random variables, each of which has a probability $p$ of being 1.

During the training process, we select one of the above mask methods to generate two perturbed embedding $\hat{\mathbf{E}}_1$ and $\hat{\mathbf{E}}_2$, where  $\hat{\mathbf{E}}_1=\mathrm{g}(\mathbf{E})$ and  $\hat{\mathbf{E}}_2=\mathrm{g}(\mathbf{E})$ in Figure \ref{fig:model_cl4ctr}(b). More analyses about the effectiveness of different mask methods are showed in Section \ref{sec:abl3}. 

\subsubsection{Feature Interaction Encoder}
We utilize a shared FI encoder to extract feature interaction information from the two perturbed embeddings $\mathbf{\hat{E}_1}$ and $\mathbf{\hat{E}_2}$ as follows:
\begin{align}
\textstyle
    h_{1} = FI_{cl}(\mathbf{\hat{E}_1}), h_{2} = FI_{cl}(\mathbf{\hat{E}_2}).
\end{align}
$FI_{cl}(\cdot)$ represents FI encoder function, and $h_1$, $h_2$ are two compressed representations generated from two perturbed embeddings. 

Notably, any FI encoder can be deployed in our CL4CTR, such as cross-network~\cite{wang2020dcn}, self-attention~\cite{song2019autoint}, and bi-interaction~\cite{he2017neural}, as described in Section \ref{sec:fi_cncoder}. Specifically, we select the Transformer layer~\cite{vaswani2017attention} as our primary FI encoder, which is widely used to extract vector-level relationships between features~\cite{xie2020CL4SREC,song2019autoint,chen2021dcap}.

Additionally, we find that the dimensions of compressed representations ($h_{1}$, $h_{2}$) generated by some FI encoders (e.g., cross network~\cite{wang2020dcn}, PIN~\cite{qu2018product}) could be huge, e.g., over thousands when field F is large, which produce adversely impacts on the training stability. Hence, we utilize a projection function to reduce the dimensions of representations from FI encoder to D as follows:
\begin{align}
\textstyle
    \hat{h}_{1}=p_1(h_{1}), \hat{h}_{2}=p_2(h_{2}).
\end{align}
The projection function $p(\cdot)$ is a single layer MLP.

\subsubsection{Contrastive Loss Function}
Finally, a contrastive loss function is applied to minimize the expected distance between the above two perturbed representations as follows:
\begin{align}
\textstyle
\label{equ:cl_loss}
    \mathcal{L}_{cl} = \frac{1}{B}\sum_{i=1}^{B}
    \left\| 
        \hat{h}_{i,1} - \hat{h}_{i,2}
    \right\|_{2}^{2}.
\end{align}
B is the batch size and  $||\cdot||_2^2$ denotes the $\ell_2$ distance. 

\subsection{Feature Alignment and Field Uniformity}
To ensure low frequency features and high-frequency features be trained equally, a naive way is to increase the frequency of low frequency features or reduce the frequency of high-frequency features during training. Inspired by previous works~\cite{wang2020understanding,wang2021understanding2} in other areas (CV, NLP), which can achieve similar goal by introducing two critical properties, named the alignment and uniformity constraints, but they need to construct positive and negative sample pairs to optimize the two constraints. In CTR prediction tasks, we find that features in the same field are analogous to positive sample pairs, and features of different fields are analogous to negative sample pairs. Thus, we propose two new properties for contrastive learning in CTR prediction, named feature alignment and field uniformity, which can regularize feature representations during training process. Specifically, feature alignment pulls the representations of features from the same field to be as close as possible. In contrast, field uniformity pushes representations of features from different fields to be as distant as possible. 

\subsubsection{Feature Alignment}
Firstly, we introduce the feature alignment constraint, which aims to minimize the distance between features from the same field. Intuitively, by adding a feature alignment constraint, the representations of features in the same field should be more closely distributed in the low-dimensional space. Formally, the loss function of feature alignment is as follows:
\begin{align}
\label{equ:align_loss}
    \mathcal{L}_{a} = \sum_{f=1}^{F} {\sum_{\mathbf{e_i},\mathbf{e_j}\in\mathbf{E}_{f}}
        {\lVert 
            \mathbf{e_i} - \mathbf{e_j}
        \rVert}_2^2,
    }
\end{align}
where $\mathbf{e_i}$ and $\mathbf{e_j}$ are two features from the same field, and $\mathbf{E}_{f}$ is the subset features of field $f$. 

\subsubsection{Field Uniformity}
The relationships among different fields have not been extensively studied in existing CTR prediction methods. For instance, FFM~\cite{juan2016field} learns field-aware representation for each feature, and NON~\cite{luo2020network} extracts intra-field information, but their techniques cannot be directly applied in contrastive learning. Differently, we introduce field uniformity to optimize feature representation directly, which minimizes the similarity between features belonging to different fields. The loss function of field uniformity is formally defined as follows:
\begin{align}
\label{equ:u_loss}
    \mathcal{L}_{u} = \sum_{\substack{\mathbf{e_i} \in \mathbf{E}_{f} \\ 1<= f <= F} }
        {
            \sum_{\mathbf{e_j}\in(\mathbb{E}-\mathbf{E}_{f})}  {sim(\mathbf{e_i},\mathbf{e_j})}
        }.
\end{align}
Similar to the other approaches~\cite{yu2022self,zhou2021selfcf}, we use cosine similarity to regularize negative sample pairs, i.e., $sim(\mathbf{e_i},\mathbf{e_j}) = {\mathbf{e_i}^T\mathbf{e_j}}/{\lVert{\mathbf{e_i}}\rVert\lVert{\mathbf{e_j}}\rVert}$. Other similarity functions can also be used here. $\mathbb{E}-\mathbf{E}_{f}$ contains all features except those from field $f$. 

In both feature alignment and field uniformity constraints, we find that low frequency features and high frequency features have equal chances to be considered. Therefore, the suboptimal representation issue for low frequency features can be largely alleviated when the two constraints are introduced during training.

\subsection{Multi-task Training}
To integrate the CL4CTR framework into the scenarios of CTR prediction, we adopt a multi-task training strategy to jointly optimize these three auxiliary SSL losses and the original CTR prediction loss in an end-to-end manner. Thus the final objective function can be formulated as follows:
\begin{align}
\label{equ:loss}
    \mathcal{L}_{total} = \mathcal{L}_{ctr}+ \alpha \cdot \mathcal{L}_{cl} + 
    \beta \cdot ( \mathcal{L}_{a} + \mathcal{L}_{u}),
\end{align}
where $\alpha$ and $\beta$ are the hyper-parameters to control the strengths of contrastive loss and feature alignment and field uniformity loss. 

\section{Experiments}
\subsection{Experimental Setup}

\begin{table}[t]
    \centering
    \caption{Dataset statistics.}
    \scalebox{0.8}{
    \begin{tabular}{c|cccc|cc}
    \hline
    \hline
   Datasets &Positive& \#Training & \#Validation & \#Test  & \#Features &\#Fields \\ \hline

   Frappe &33\%  &202K & 58K & 29K & 5K &10\\ 
   ML-tag  &33\%  &1,404K  &401K  &201K &90K  &3\\ 
   ML-1M & 57.5\%  &800K  &100K  &100K &10K &5\\
   SafeDriver &3.64\%  &476K &59K  & 59K   &600 &57\\
    \hline
    \hline
    \end{tabular}
    }
    \label{Tab.dataset}
\end{table}

\subsubsection{Datasets}
We evaluate CL4CTR on four popular datasets: \textbf{Frappe}\footnote{https://www.baltrunas.info/context-aware/frappe}~\cite{he2017neural}, \textbf{ML-tag}\footnote{https://grouplens.org/datasets/movielens/}~\cite{he2017neural},
\textbf{SafeDriver}\footnote{\url{https://www.kaggle.com/c/porto-seguro-safe-driver-prediction}}~\cite{huang2020gatenet} 
and \textbf{ML-1M}\footnote{https://grouplens.org/datasets/movielens/1m/}~\cite{chen2021dcap}. 
The statistics of the four datasets are presented in Table \ref{Tab.dataset}. NFM~\cite{he2017neural} and AFM~\cite{xiao2017attentional} have strictly split Frappe and ML-tag to training, validation, and testing by 7:2:1, and we directly follow their settings. For SafeDriver and ML-1M, following ~\cite{huang2020gatenet} and ~\cite{chen2021dcap}, we randomly split the instances by 8:1:1. Detailed descriptions of those datasets can be found in the links or references.

% ~\cite{meng2021general}
\subsubsection{Compared Methods}
To evaluate the proposed CL4CTR framework, we compare its performance with four classes of representative CTR methods~\cite{meng2021general}. 
1) First-order method that is a weighted sum of raw features, including \textbf{LR};
2) FM-based methods that consider second-order FI, including  \textbf{FM}~\cite{rendle2012factorization}, 
% \textbf{FFM}~\cite{juan2016field}, 
\textbf{FwFM}~\cite{pan2018field}, 
\textbf{IFM}~\cite{yu2019input}, and
\textbf{FmFM}~\cite{sun2021fm2};
3) Approaches that model higher-order FI, including 
% \textbf{HOFM}~\cite{blondel2016higher}
\textbf{CrossNet}~\cite{wang2017deep}, 
% \textbf{NFM}~\cite{he2017neural},
\textbf{IPNN}~\cite{qu2018product}, 
\textbf{OPNN}~\cite{qu2018product},
% \textbf{CIN}~\cite{lian2018xdeepfm},
\textbf{FINT}~\cite{zhao2022fint}, and 
\textbf{DCAP}~\cite{chen2021dcap};
4) Ensemble methods or multi-tower structures, including 
\textbf{WDL}~\cite{cheng2016wide}, 
\textbf{DCN}~\cite{wang2017deep}, 
\textbf{DeepFM}~\cite{guo2017deepfm}, 
\textbf{xDeepFM}~\cite{lian2018xdeepfm}, 
\textbf{FiBiNET}~\cite{huang2019fibinet},  
\textbf{AutoInt+}~\cite{song2019autoint}, 
\textbf{AFN+}~\cite{cheng2020adaptive}, 
% \textbf{NON}~\cite{luo2020network}, 
\textbf{TFNET}~\cite{wu2020tfnet}, 
\textbf{FED}~\cite{zhao2020fed_dimension}, and 
\textbf{DCN-V2}~\cite{wang2020dcn}.
The proposed CL4CTR framework is model-agnostic. For simplicity, a base model \textbf{\textit{M}} equipped with CL4CTR is represented as $CL4CTR_\textbf{\textit{M}}$. We choose FM~\cite{rendle2012factorization} as the basic model to verify the effectiveness of CL4CTR, which only models second-order FI and has no additional parameters except for feature representations. \textbf{Therefore, the performance boost of $CL4CTR_{FM}$ directly reflects the quality of the feature representations.} CL4CTR only helps the base CTR models training and does not add any operation or parameter to the inference process.

\subsubsection{Evaluation Metrics}
To evaluate the performance of all methods, we adopt the commonly-used \textbf{AUC} (Area Under ROC) and \textbf{Logloss} (cross entropy) as the metrics. Notably, a slightly higher AUC or a lower Logloss at \textbf{0.001}-level can be considered significant for CTR prediction ~\cite{chen2021enhancing,luo2020network, wang2020dcn, huang2019fibinet,lian2018xdeepfm}.

\subsubsection{Implementation Details.} For fair comparisons, we implement all the models with Pytorch\footnote{ The code of CL4CTR is available here: \url{ https://github.com/cl4ctr/cl4ctr}} and optimize all models with Adam. The embedding size is set to 64 for Frappe and ML-tag and 20 for ML-1M and SafeDriver, respectively. Meanwhile, the batch size is fixed to 1024. the learning rate is 0.01 for SafeDriver and 0.001 for other datasets. As for the models including DNN in the prediction layer, we adopt the same structure $\{$400,400,400,1$\}$. All the activation functions are ReLU, and the dropout rate is 0.5. We perform early stopping according to AUC on the validation set to avoid overfitting. We also implement the Reduce-LR-On-Plateau scheduler to reduce the learning rate by a factor of 10 when the given metric stops improving within four continuous epochs. Each experiment is repeated 5 times, and the average results are reported. In CL4CTR, $FI_{cl}(\cdot)$ adopts three transformer layers. And we use hyper-parameters: $\alpha$=1, $\beta$=0.01 in the final loss function. 
% \footnote{\url{https://pytorch.org/docs/stable/optim.html}}

\begin{table*}[t]
    \setlength{\abovecaptionskip}{0.2cm}
    \setlength{\belowcaptionskip}{-0.2cm}
\centering
\caption{Overall accuracy comparison in the four datasets. $\Delta{AUC}$ and $\Delta{Logloss}$ indicate averaged performance boost compared with DCN-V2. \textit{RelaImp} denotes the relative improvements compared with the strongest baseline. Bold scores are the best performance, while underlined scores are the second best. Improvements over baselines are statistically significant with p<0.01.} 
\label{tab:all}
\scalebox{0.95}{
\begin{tabular}{cc|cc|cc|cc|cc|cc} 
\hline \hline

\multirow{2}{*}{\begin{tabular}[c]{@{}c@{}}Model\\Class\end{tabular}}
& Datasets
& \multicolumn{2}{c|}{Frappe} 
& \multicolumn{2}{c|}{ML-tag} 
&\multicolumn{2}{c|}{ML-1M}
& \multicolumn{2}{c}{SafeDriver}
&\multirow{2}{*}{\begin{tabular}[c]{@{}c@{}}$\Delta{AUC} $\\ $\uparrow$ \end{tabular}} &
\multirow{2}{*}{\begin{tabular}[c]{@{}c@{}}$\Delta{Logloss} $\\$\downarrow$\end{tabular}} \\

\cline{2-10}
& Model& AUC& Logloss& AUC& Logloss& AUC& Logloss& AUC& \multicolumn{1}{c}{Logloss} & \multicolumn{1}{l}{}& \multicolumn{1}{l}{}\\ 
\hline
\multirow{1}{*}
{\begin{tabular}[c]{@{}c@{}}
First-order\end{tabular}}    
& LR         & 0.9331 & 0.2894  & 0.9348 & 0.2960   & 0.7899       & 0.5417  & 0.6244     & 0.1622  & -3.35\%  & 0.0572  \\
\hline
\multirow{4}{*}{Second-Order} 
& FM         & 0.9746 & 0.1856  & 0.9488 & 0.2595  & 0.8023       & 0.5332  & 0.6301     & 0.1538  & -1.22\%  & 0.0179  \\
& FwFM    & 0.9756 & 0.1784  & 0.9582 & 0.2531  & 0.8046       & 0.5281  & 0.6335     & 0.1532  & -0.74\%  & 0.0131  \\
& IFM        & 0.9771 & 0.1581  & 0.9515 & 0.2497  & 0.8080        & 0.5286  & 0.6353     & 0.1526  & -0.70\%  & 0.0071  \\
& FmFM       & 0.9801 & 0.1682  & 0.9552 & 0.2493  & 0.8093       & 0.5264  & 0.6378     & 0.1518  & -0.39\%  & 0.0088  \\
\hline
\multirow{5}{*}{High-Order}   
% & HOFM       & 0.9759 & 0.1835  & 0.9478 & 0.2656  & 0.8025       & 0.5317  & 0.6366     & 0.1516  & -0.95\%  & 0.0180  \\
& CrossNet   & 0.9800   & 0.1658  & 0.9549 & 0.2480   & 0.8114       & 0.5218  & 0.6336     & 0.1517  & -0.50\%  & 0.0067  \\
& IPNN       & \underline{0.9809} & 0.1604  & 0.9607 & 0.2295  & 0.8110        & 0.5190   & 0.6373     & 0.1521  & -0.19\%  & 0.0001  \\
& OPNN       & 0.9799 & 0.1683  & 0.9599 & 0.2421  & 0.8112       & 0.5185  & 0.6375     & 0.1519  & -0.22\%  & 0.0051  \\
& FINT       & 0.9807 & \underline{0.1578}  & 0.9557 & 0.2430   & 0.8123       & 0.5192  & 0.6349     & 0.1522  & -0.38\%  & 0.0029  \\
& DCAP       & 0.9801 & 0.1612  & 0.9560  & 0.2428  & 0.8130        & 0.5171  & 0.6390      & 0.1512  & -0.20\%  & 0.0030   \\
\hline
\multirow{10}{*}{Ensemble}    
& WDL        & 0.9770  & 0.1783  & 0.9599 & 0.2660   & 0.8093       & 0.5226  & 0.6353     & 0.1525  & -0.44\%  & 0.0110  \\
& DCN        & 0.9788 & 0.1621  & 0.9550  & 0.2472  & 0.8125       & 0.5175  & 0.6379     & 0.1514  & -0.32\%  & 0.0044  \\
& DeepFM     & 0.9780  & 0.1732  & 0.9586 & 0.2551  & 0.8061       & 0.5259  & 0.6318     & 0.1529  & -0.69\%  & 0.0117  \\
& xDeepFM   & 0.9799 & 0.1750   & 0.9604 & 0.2472  & 0.8082       & 0.5244  & 0.6403     & 0.1515  & -0.19\%  & 0.0094  \\
& FiBiNET    & 0.9793 & 0.1707  & 0.9548 & 0.2532  & 0.8032       & 0.5313  & 0.6391     & \underline{0.1505}  & -0.56\%  & 0.0113  \\
& AutoInt+   & 0.9783 & 0.1762  & 0.9535 & 0.2562  & 0.8099       & 0.5219  & 0.6310      & 0.1516  & -0.73\%  & 0.0114  \\
& AFN+       & 0.9786 & 0.1637  & 0.9561 & 0.2468  & 0.8041       & 0.5304  & 0.6374     & 0.1517  & -0.58\%  & 0.0080   \\
% & NON        & 0.9806 & 0.1605  & 0.9578 & 0.2556  & 0.8110        & 0.5193  & 0.6389     & 0.1528  & -0.21\%  & 0.0069  \\
& TFNet      & 0.9798 & 0.1708  & 0.9527 & 0.2551  & 0.8099       & 0.5212  & 0.6387     & 0.1533  & -0.41\%  & 0.0100  \\
& FED        & 0.9791 & 0.1606  & 0.9557 & 0.2465  & 0.8128       & 0.5184  & 0.6369     & 0.1534  & -0.33\%  & 0.0046  \\
& DCN-V2     & 0.9803 & 0.1595  & \underline{0.9610} & \underline{0.2330}   & \underline{0.8132}       & \underline{0.5169}  & \underline{0.6406}     & 0.1510   & -  & -       \\
\hline
\multirow{2}{*}{\begin{tabular}[c]{@{}c@{}}Ours\end{tabular}} 
&$CL4CTR_{FM}$ &\textbf{0.9822}	&\textbf{0.1324}	&\textbf{0.9621}	&\textbf{0.2102}	&\textbf{0.8164}	&\textbf{0.5136}	&\textbf{0.6449}	&\textbf{0.1483}& 0.34\% & -0.0140\\
\multirow{2}{*}{\begin{tabular}[c]{@{}c@{}}\end{tabular}} 
& \textit{RelaImp} & 0.13\% & 16.10\% & 0.11\% & 8.41\%  & 0.39\%       & 0.64\%  & 0.67\% & 1.46\%  & -  & - \\
\hline
\hline
\end{tabular}
}
\end{table*}

\subsection{Overall Comparison}

In this section, we compare the performances of $CL4CTR_{FM}$ with the state-of-the-art (SOTA) CTR prediction models. Table \ref{tab:all} shows the experimental results of all compared models over four datasets.

It can be observed that LR and FM achieve the worst performance compared with other baselines, which indicates that shallow models are insufficient for CTR prediction. Other FM-based models improve FM by introducing field importance mechanism (e.g., FwFM~\cite{pan2018field} and IFM~\cite{yu2019input}) or deploying a novel field-pair matrix approach (e.g., FmFM~\cite{sun2021fm2}). Generally, deep-learning based models (e.g., DeepFM~\cite{guo2017deepfm}, DCN~\cite{wang2017deep}, DCN-V2~\cite{wang2020dcn}), which combine high-order feature interactions with well-designed feature interaction structures, achieve better performance than FM.

$CL4CTR_{FM}$ consistently performs better than all baselines on all datasets. Furthermore, $CL4CTR_{FM}$ significantly outperforms the strongest baseline DCN-V2~\cite{wang2020dcn} by 0.13\%, 0.11\%, 0.39\% and 0.67\% in terms of AUC (16.10\%, 8.41\%, 0.64\% and 1.46\% in terms of Logloss) on Frappe, ML-tag, ML-1M, and SafeDriver respectively. Additionally, we find that the improvements on Logloss are more remarkable than those on AUC, indicating that CL4CTR enables us to predict the true click probability effectively. Meanwhile, $CL4CTR_{FM}$ shows strong generalization ability on all datasets, where Table  \ref{tab:all} shows the averaged performance boost ($\Delta$AUC and $\Delta$Logloss). Notably, most SOTA CTR prediction models design complex networks to produce advanced feature representations and useful feature interactions to improve the performance. However, our CL4CTR only helps FM to learn accurate feature representations from the embedding layer with contrastive learning instead of introducing extra modules. The improvement of our CL4CTR verifies the necessity of learning accurate feature representations in CTR prediction tasks.

\subsection{Ablation Study}
\label{sec:abl3}
\subsubsection{Compatibility Analysis}
\label{sec:comp}
To verify the compatibility of CL4CTR, we deploy it into other SOTA models, such as DeepFM~\cite{guo2017deepfm}, Autoint+~\cite{song2019autoint}, and DCN-V2~\cite{wang2020dcn}. The results are shown in Table \ref{tab:abl_compatibility}.  

Firstly, learning feature representation with CL4CTR can significantly improve the performance of CTR prediction. Applied with CL4CTR, the performance of base models is remarkably boosted, which confirms our hypothesis of improving the performance of CTR prediction models by improving the quality of the feature representations and demonstrates the effectiveness of CL4CTR. In addition, the experimental results show that learning high-quality feature representations is at least as important as designing advanced FI techniques. Modeling complex feature interactions can improve the performance of CTR models when these models leverage supervised signal for training, which explains why FI-based models outperform FM. However, after introducing self-supervised signals into CTR models for learning high-quality feature representations, FM can achieve the best performance compared with other models deployed with CL4CTR. The possible reason is that both supervised and self-supervised learning are directly and only optimizing the parameters in feature representations in FM without disturbing by other parameters. 

\begin{table}
    \setlength{\abovecaptionskip}{0.2cm}
    \setlength{\belowcaptionskip}{-0.2cm}
\centering
\caption{Compatibility study of CL4CTR.}
\label{tab:abl_compatibility}
\scalebox{0.85}{
\begin{tabular}{ccccccc} 
\toprule
\multicolumn{1}{c}{\multirow{2}{*}{Model}} & \multicolumn{2}{c}{Frappe} & \multicolumn{2}{c}{ML-1M}  & \multicolumn{2}{c}{SafeDriver}  \\ 
\cline{2-7}
\multicolumn{1}{c}{}  & AUC & Logloss& AUC & Logloss& AUC & Logloss   \\ 
\hline
FM    
& 0.9746   & 0.1856   & 0.8023   & 0.5332   &0.6244&0.1622\\
$CL4CTR_{FM}$ 
& \textbf{0.9822} & \textbf{0.1324} & \textbf{0.8164} & \textbf{0.5136} &\textbf{0.6449}&\textbf{0.1483}\\ 
\hline
FwFM  & 0.9756   & 0.1784   & 0.8046   & 0.5281   &0.6335&0.1532\\
$CL4CTR_{FwFM}$   & \textbf{0.9815} & \textbf{0.1532} & \textbf{0.8118} & \textbf{0.5192} &\textbf{0.6421}&\textbf{0.1487}\\ 
\hline
DeepFM
& 0.9780   & 0.1732   & 0.8061   & 0.5259 &0.6318&0.1529\\
$CL4CTR_{DeepFM}$
& \textbf{0.9813}   & \textbf{0.1677}   & \textbf{0.8113}   & \textbf{0.5194}   &\textbf{0.6381} &\textbf{0.1504}\\ 
\hline
Autoint+  
& 0.9783   & 0.1762   & 0.8099   & 0.5219   &0.6310 &0.1516\\
$CL4CTR_{Autoint+}$ 
& \textbf{0.9802}& \textbf{0.1684} & \textbf{0.8122}   & \textbf{0.5174}   &\textbf{0.6402}&\textbf{0.1506}\\ 
\hline
DCN   
& 0.9788   & 0.1621   & 0.8125   & 0.5170   &0.6379&0.1514\\
$CL4CTR_{DCN}$
& \textbf{0.9808} & \textbf{0.1566}   & \textbf{0.8164}   & \textbf{0.5125}   &\textbf{0.6415}&\textbf{0.1494}\\ 
\hline
DCN-V2
& 0.9803   & 0.1595   & 0.8132   & 0.5169   &0.6406 &0.1510\\
$CL4CTR_{DCN-V2}$
& \textbf{0.9812}   & \textbf{0.1549}   & \textbf{0.8144}   & \textbf{0.5153}   &\textbf{0.6411}&\textbf{0.1497}\\ 
\bottomrule
\end{tabular}}
\end{table}

\subsubsection{Data Augmentation Approaches}
\label{sec:augmen}
To verify the effectiveness of our proposed data augmentation methods, we change the augmentation methods in the contrastive module and fix other settings for a fair comparison. Furthermore, we select different baseline models and deploy CL4CTR into them to compare their performance under this setting. Table \ref{tab:abl_mask} shows the experimental results. 

The random mask method achieves the best performance in most cases. We think the random mask is more moderated than feature mask and dimension mask because it omits element information. Additionally, the FwFM model achieves the best performance with the feature mask method on Frappe; in contrast, it achieves the best performance with the dimension mask method on SafeDriver, demonstrating that our proposed augmentation methods are effective and can be used in different baseline models and datasets.

\begin{table}
    \setlength{\abovecaptionskip}{0.2cm}
    \setlength{\belowcaptionskip}{-0.2cm}
\caption{Impact of data augmentation methods.}
\label{tab:abl_mask}
\centering
\scalebox{0.85}{
\begin{tabular}{cccccc} 
\toprule
\multirow{2}{*}{\begin{tabular}[c]{@{}c@{}}Base \\model\end{tabular}} & \multirow{2}{*}{Variants} & \multicolumn{2}{c}{Frappe} & \multicolumn{2}{c}{SafeDriver}  \\ 
\cline{3-6}
& & AUC & Logloss & AUC    & Logloss\\ 
\hline
\multirow{4}{*}{FM}    
& Base &0.9746&0.1856   &0.6244  &0.1622\\
& Random    &\textbf{0.9822}&\textbf{0.1324}   &\textbf{0.6449}  &\textbf{0.1483} \\
& Feature   &0.9814&0.1328  &0.6303  &0.1539\\
& Dimension &0.9816&0.1334  &0.6404  &0.1505  \\ 
\hline
\multirow{4}{*}{FwFM} 
& Base &0.9756&0.1784 &0.6335 &0.1532\\
& Random    &0.9815  &0.1532 & 0.6421 &0.1487\\
& Feature   &\textbf{0.9822}&\textbf{0.1513}  &0.6384  &0.1483\\
& Dimension &0.9811&0.1465  &\textbf{0.6455} &\textbf{0.1508}\\
\hline
\multirow{4}{*}{DeepFM} 

& Base      &0.9780&0.1817 &0.6318 &0.1529\\
& Random    &\textbf{0.9813} &\textbf{0.1677} & \textbf{0.6381} &\textbf{0.1504}\\
& Feature   &0.9798 &0.1750  &0.6341  &0.1522\\
& Dimension &0.9804 &0.1697  &0.6353  &0.1514 \\
\hline
\multirow{4}{*}{DCN} 
& Base &0.9788&0.1611 &0.6379 &0.1514\\
& Random    &\textbf{0.9808} &\textbf{0.1566} & \textbf{0.6415} &\textbf{0.1494}\\
& Feature   &0.9804 &0.1601  &0.6409  &0.1508\\
& Dimension &0.9803 &0.1573  &0.6411  &0.1504 \\
\bottomrule
\end{tabular}}
\end{table}

\begin{table}
    \setlength{\abovecaptionskip}{0.2cm}
    \setlength{\belowcaptionskip}{-0.2cm}
\caption{Impact of different FI encoder $FI_{cl}(\cdot)$.}
\label{tab:abl_contrastive}
\centering
\scalebox{0.85}{
\begin{tabular}{cccccc}
\hline
\multirow{2}{*}{\begin{tabular}[c]{@{}c@{}}Base \\model\end{tabular}} & \multicolumn{1}{c}{\multirow{2}{*}{\begin{tabular}[c]{@{}c@{}}FI \\Encoder\end{tabular}}} 
& \multicolumn{2}{c}{Frappe}& \multicolumn{2}{c}{ML-1M}   \\ 
\cline{3-6}
& \multicolumn{1}{c}{} & AUC & Logloss& AUC& Logloss \\ 
\hline
\multirow{4}{*}{FM}    
& Base         &0.9746    &0.1856    &0.8023&0.5332\\
& DNN          &0.9804	&0.1404 & \textbf{0.8177}	&\textbf{0.5123} \\
& Transformer  &\textbf{0.9822}&\textbf{0.1324} & 0.8164    &0.5136 \\
& CrossNet2    &0.9801&0.1438&0.8170	&0.5143\\
\hline
\multirow{4}{*}{FwFM}   
& Base         &0.9756&0.1784 &0.8046  &0.5281\\
& DNN          &0.9809&\textbf{0.1504}&0.8064	&0.5264\\
& Transformer  &0.9815&0.1532    &\textbf{0.8118}&\textbf{0.5192}  \\
& CrossNet2    &\textbf{0.9822}    &0.1675  &0.8102	&0.5231 \\
\hline
\multirow{4}{*}{DeepFM} 
& Base         &0.9780&0.1732&0.8061&0.5259\\
& DNN          &0.9804&0.1710&0.8101&0.5206\\
& Transformer  &\textbf{0.9813} &\textbf{0.1704}&\textbf{0.8113}&\textbf{0.5194}\\
& CrossNet2    &0.9791&0.1719&0.8109&0.5202\\   
\hline
\multirow{4}{*}{DCN-V2}  
& Base          &0.9803&0.1595&0.8132&0.5169\\
& DNN           &0.9807&0.1573&\textbf{0.8151}&\textbf{0.5144}\\
& Transformer   &\textbf{0.9812} &\textbf{0.1549} &0.8144&0.5153\\
& CrossNet2     &0.9804&0.1588&0.8141&0.5155\\ 
\hline
\end{tabular}
}
\end{table}

\subsubsection{FI Encoder}
In the contrastive module, the FI encoder also affects the performance of CL4CTR, as different structures of the FI encoder extract different information. For instance, transformer layer can model high-order feature interactions in feature-level~\cite{song2019autoint,chen2021dcap} explicitly, however, CrossNet2~\cite{wang2020dcn} focuses on modelling bounded-degree feature interactions in element-level explicitly. DNN is a common and widely used structure in most CTR models for modeling bit-level feature relationships implicitly. We select the above three representative structures as FI encoders and verify their performance. Notably, we adopt three layers structure as reported in their paper. Table \ref{tab:abl_contrastive} shows the experimental results.

It can be observed that CL4CTR can consistently improve the performance of these baseline models with different FI encoders. In addition, since different FI encoders extract different information based on specific base models and datasets, the performance of these models is different. However, CL4CTR with transformer layers achieves the best performance in most cases since the transformer layer is a more effective solution than others.

\subsubsection{Loss Function}
In this section, we evaluate the effectiveness of self-supervised learning signals ( i.e., $\mathcal{L}_{cl}$, $\mathcal{L}_a$, $\mathcal{L}_u$ ) by eliminating them from CL4CTR respectively. We still regard $\mathcal{L}_a$ and $\mathcal{L}_u$ as a whole part. The experimental results are shown in Table \ref{tab:abl_loss}.

Firstly, we can find that each self-supervised learning signal deployed in baseline models can improve their performance. In addition, by comparing the contrastive loss and alignment\&uniformity constraints individually, we conclude that they play different roles in different datasets and baseline models. Specifically, FM with $\mathcal{L}_a$ and $\mathcal{L}_u$ perform better than FM with $\mathcal{L}_{cl}$; in contrast, DCN with $\mathcal{L}_a$ and $\mathcal{L}_u$ perform worse than DCN with $\mathcal{L}_{cl}$, which verifies our hypothesis. Furthermore, all experiments achieve the best performance when $\mathcal{L}_{cl}$, $\mathcal{L}_a$, and $\mathcal{L}_u$ are deployed simultaneously.

Compared with individual SSL signals, we find that training with all of them can always achieve the best performance. Furthermore, adopting $\mathcal{L}_a$ and $\mathcal{L}_u$ in FM consistently outperforms adopting $\mathcal{L}_{cl}$ in FM on two datasets evaluated by Logloss, which indicates that inducing feature alignment and field uniformity into CTR prediction models enables us to predict probabilities closer to the true label.

\begin{table}
    \setlength{\abovecaptionskip}{0.2cm}
    \setlength{\belowcaptionskip}{-0.2cm}
\caption{Impact of SSL signals in the loss function.}
\label{tab:abl_loss}
\centering
\scalebox{0.85}{
\begin{tabular}{cclccc} 
\toprule
\multirow{2}{*}{\begin{tabular}[c]{@{}c@{}}Model\end{tabular}} & \multirow{2}{*}{Loss Function} & \multicolumn{2}{c}{Frappe} & \multicolumn{2}{c}{ML-1M}  \\ 
\cline{3-6}
& & AUC & Logloss & AUC    & Logloss\\ 
\hline
\multirow{4}{*}{FM}
& $\mathcal{L}_{ctr}$ &0.9746&0.1856 &0.8023&0.5332\\
& + $\mathcal{L}_{cl}$ &0.9794  &0.1485 &0.8102 &0.5230\\
& + $(\mathcal{L}_{a} +\mathcal{L}_{u})$  &\underline{0.9812}&\underline{0.1455}&\underline{0.8139} &\underline{0.5175}\\
& + $\mathcal{L}_{cl}$ + $(\mathcal{L}_{a} +\mathcal{L}_{u})$  &\textbf{0.9822}&\textbf{0.1324} &\textbf{0.8164} &\textbf{0.5136} \\ 
\hline

\multirow{4}{*}{FwFM}   
& $\mathcal{L}_{ctr}$ &0.9756&0.1784 &0.8046 &0.5281\\
& + $\mathcal{L}_{cl}$    &0.9785&0.1553 &\underline{0.8109} &\underline{0.5229}\\
& + $(\mathcal{L}_{a} +\mathcal{L}_{u})$ &\underline{0.9812}&\underline{0.1536} &0.8098 &0.5252\\
& + $\mathcal{L}_{cl}$ + $(\mathcal{L}_{a} +\mathcal{L}_{u})$&\textbf{0.9815}&\textbf{0.1532} &\textbf{0.8118}&\textbf{0.5192} \\ 
\hline
\multirow{4}{*}{DeepFM} 
& $\mathcal{L}_{ctr}$&0.9780 &0.1817 &0.8061 &0.5259\\
& + $\mathcal{L}_{cl}$    &\underline{0.9794} &\underline{0.1701} &0.8094 &0.5235\\
& + $(\mathcal{L}_{a} +\mathcal{L}_{u})$ &0.9784 &0.1791 &\underline{0.8103} &\underline{0.5214}\\
& + $\mathcal{L}_{cl}$ + $(\mathcal{L}_{a} +\mathcal{L}_{u})$&\textbf{0.9813} &\textbf{0.1677} &\textbf{0.8113} &\textbf{0.5194}\\ 
\hline
\multirow{4}{*}{DCN}  
& $\mathcal{L}_{ctr}$&0.9788&0.1611 &0.8125 &0.5170\\
& + $\mathcal{L}_{cl}$ &\underline{0.9802}&\underline{0.1585} &\underline{0.8138} &\underline{0.5150}\\
& + $(\mathcal{L}_{a} +\mathcal{L}_{u})$  &0.9792&0.1600 &0.8129 &0.5188\\
& + $\mathcal{L}_{cl}$ + $(\mathcal{L}_{a} +\mathcal{L}_{u})$ &\textbf{0.9808}&\textbf{0.1566} &\textbf{0.8164} &\textbf{0.5125} \\ 
\bottomrule
\end{tabular}}
\end{table}

\begin{figure}[tb]
    \setlength{\abovecaptionskip}{0.2cm}
    \setlength{\belowcaptionskip}{-0.2cm}
\centering
\includegraphics[width=0.45\textwidth]{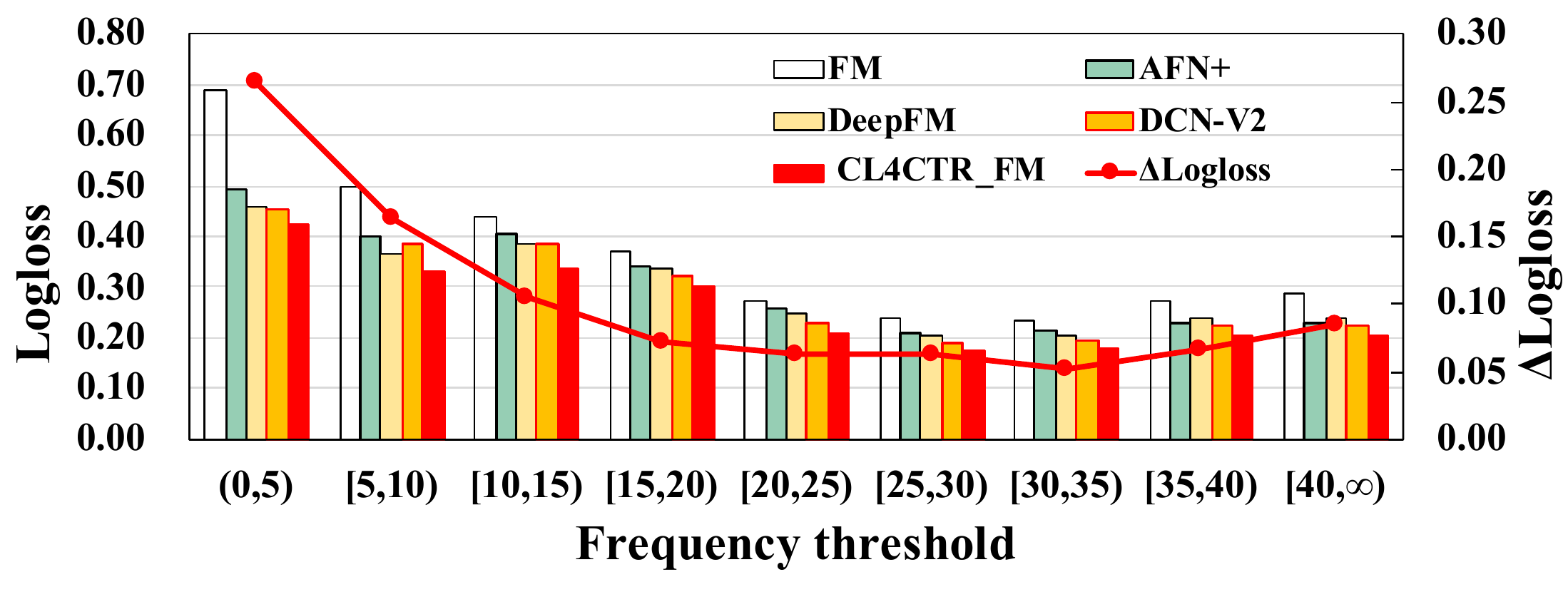}
\caption{Improvement vs. feature frequency.  }
\label{fig:case_ll}
\end{figure}

\begin{figure}[tb]
\setlength{\abovecaptionskip}{0.2cm}
\setlength{\belowcaptionskip}{-0.2cm}
\centering
\subfloat[The AUC of Frappe]{
\begin{minipage}[t]{0.52\linewidth}
\centering
\includegraphics[width=1\textwidth]{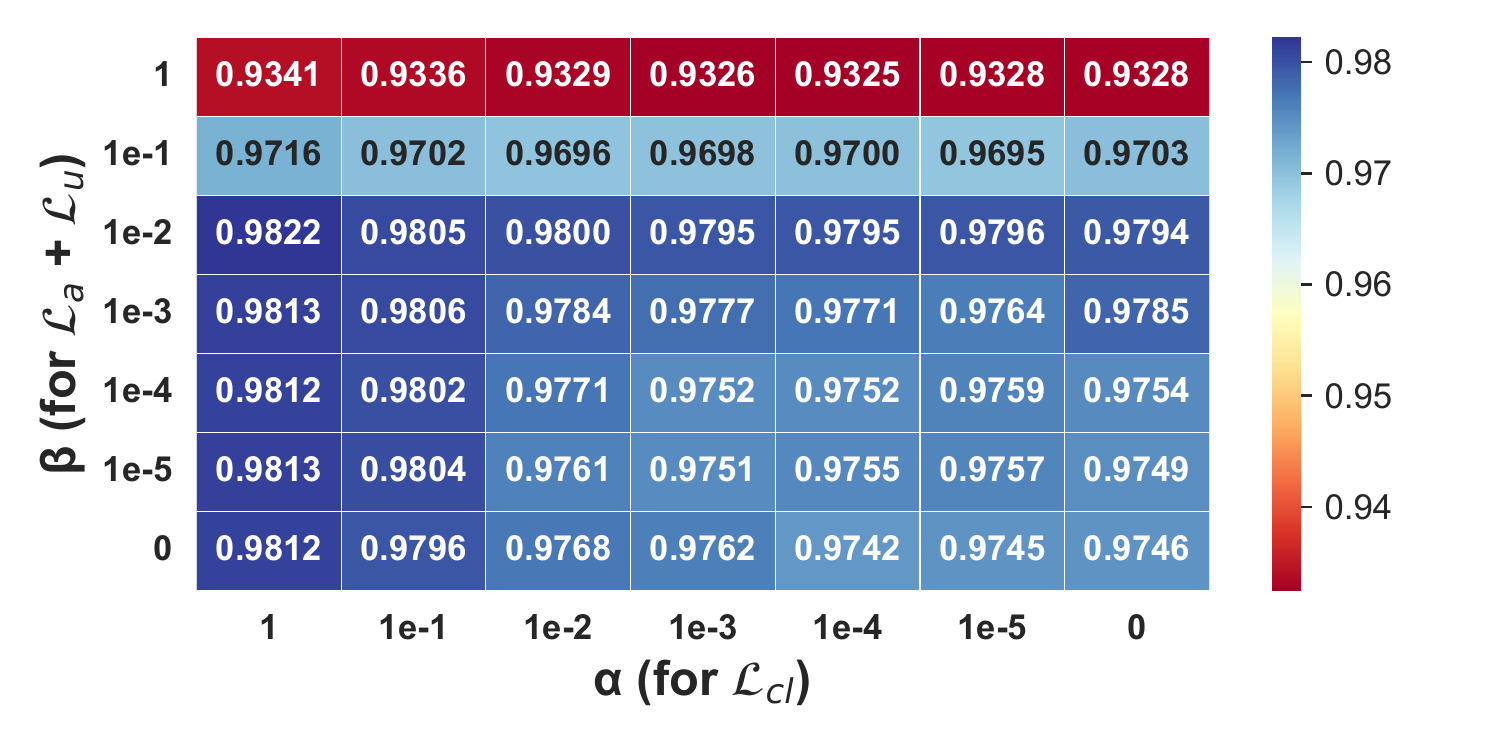}
\end{minipage}
}
\subfloat[The AUC of ML-1M]{
\begin{minipage}[t]{0.52\linewidth}
\centering
\includegraphics[width=1\textwidth]{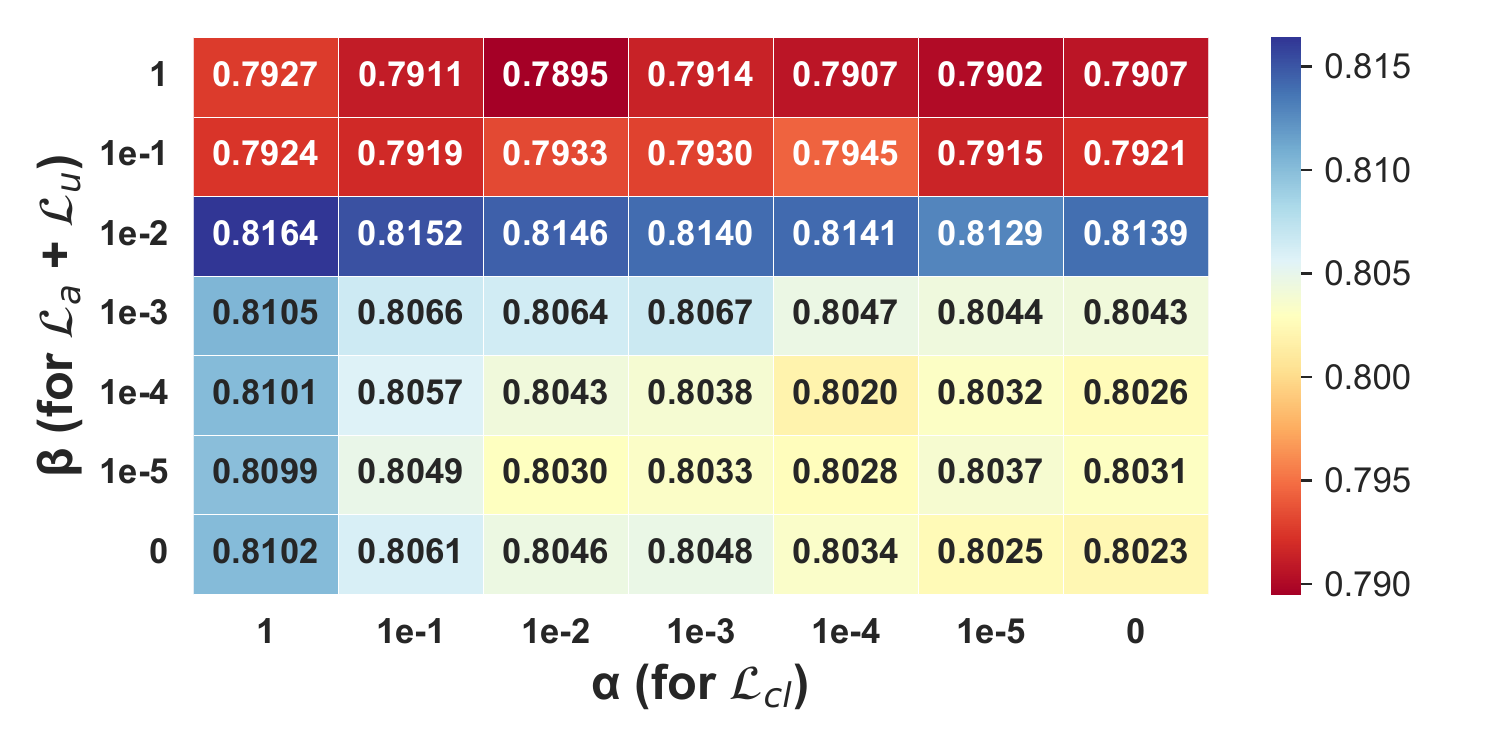}
\end{minipage}
}
\centering
\caption{Performance of $CL4CTR_{FM}$ w.r.t. different weights assigned to three SSL signals: $\alpha$ for $\mathcal{L}_c$, $\beta$ for $\mathcal{L}_a$ and $\mathcal{L}_u$.}
\label{fig:loss_weight}
\end{figure}

\subsection{Feature Frequency Analysis}
To verify the effects of feature frequency on different models, we divide the test set of ML-tag according to feature frequency and calculate the corresponding Logloss, where $\Delta Logloss$ represents the improvement over the base FM model after applying CL4CTR. Figure \ref{fig:case_ll} shows the experimental results.

Firstly, the low frequency features adversely affect the accuracy of the base model. Specifically, we show the performance of FM, three SOTA models (AFN+, DeepFM, DCN-V2), and $CL4CTR_{FM}$ in different frequency ranges. It can be observed that all models perform the worst when the input subset contains low frequency features. With the increasing of feature frequency, the performance of all models improves consistently. When the feature frequency is over 20, the performance of all models becomes stable. Figure \ref{fig:case_ll} confirms our hypothesis that only using the back-propagation to learn the  representations of low frequency features with a single supervised signal cannot achieve optimal performance. 

Secondly, CL4CTR can effectively alleviate the negative effects caused by low frequency features and keep achieving the best performance among different feature frequency ranges. By applying the alignment\&uniformity constraints,  we ensure the low frequency features can be optimized in each back-propagation process with equal chances to high frequency features. Additionally, the contrastive module can also improve the quality of the representations of all features, including both low and high frequency features.

\subsection{Hyper-parameter Analysis}
\subsubsection{Impact of the  Weights in the Loss Function}
We further investigate the impact of different weights ($\alpha$ and $\beta$ in Equation \ref{equ:loss}). We tune both $\alpha$ and $\beta$ from $\{$1, 1e-1, 1e-2, 1e-3, 1e-4, 1e-5, 0$\}$. We keep other settings fixed for fair comparison. Figure \ref{fig:loss_weight} shows the experimental results. Additionally, the trend of Logloss is consistent with AUC on those two datasets.

Overall, CL4CTR achieves the best performance when $\alpha$ is 1, and $\beta$ is 1e-2 for Frappe and ML-1M datasets. Specifically, the performance of CL4CTR deteriorates when $\alpha$ is less than 1. In addition, when $\beta$ is over 1e-2 (i.e., 1 or 1e-1), the performance of CL4CTR is significantly reduced. Meanwhile, CL4CTR performs worse with lower $\alpha$ and $\beta$ (lower right corners). 

\begin{figure}[t]
    \setlength{\abovecaptionskip}{0.1cm}
    \setlength{\belowcaptionskip}{-0.2cm}
    \centering
    \includegraphics[width=0.43\textwidth]{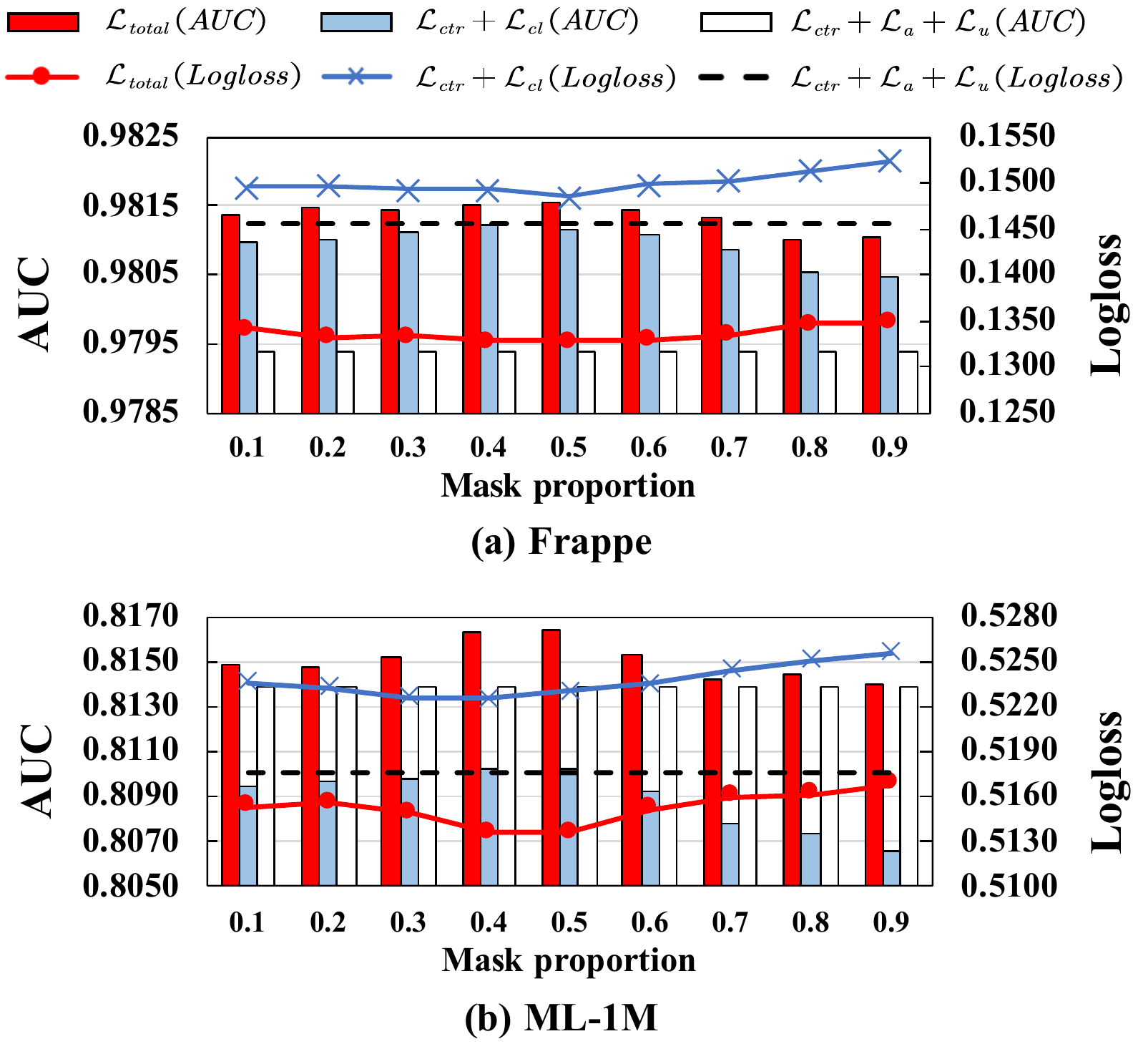}
    \caption{Impact of random mask proportion.}
\label{fig:hp_mask_ratio}
\end{figure}

\subsubsection{Mask Proportion}
We change the mask proportion $p$ within the range (0, 1) with a step size of 0.1. Note that the mask proportion is only applied in $\mathcal{L}_{cl}$. Figure \ref{fig:hp_mask_ratio} shows the results.

For models with $\mathcal{L}_{cl}$, their performance shows similar trends on Frappe and ML-1M. When the mask proportion is around 0.4 or 0.5, $CL4CTR_{FM}$ can achieve the best performance. Specifically, the model performance decreases slightly when smaller mask proportions (i.e., 0.1 to 0.3) are chosen. When mask proportion is over 0.5, the model performance decreases consistently, which is because the FI encoders only use a small percentage of information to produce valid interaction representations for calculating contrastive loss with higher mask proportions.

\begin{figure}[t]
    \setlength{\abovecaptionskip}{0.1cm}
    \setlength{\belowcaptionskip}{-0.2cm}
    \centering
    \includegraphics[width=0.50\textwidth]{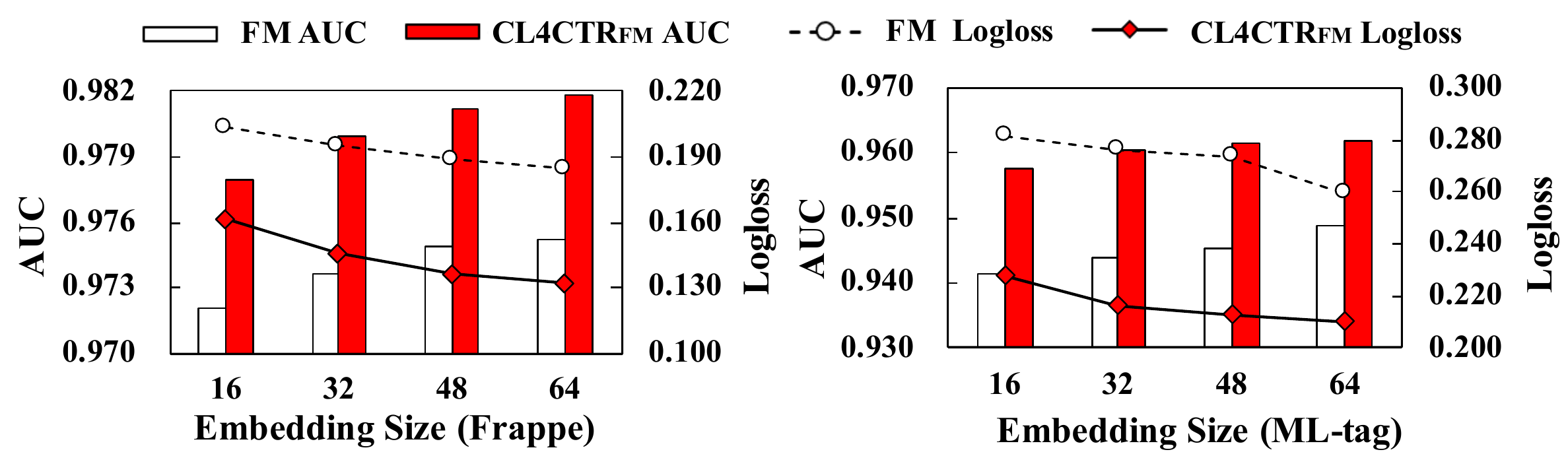}
    \caption{Impact of embedding size on FM and \textbf{$CL4CTR_{FM}$}.}
    \label{fig:hp_dim}
\end{figure}

\subsubsection{Embedding Size}
We change the embedding size from 16 to 64 with a step of 16 in the embedding layer and show the experimental results in Figure \ref{fig:hp_dim}. It can be observed that the performance of $CL4CTR_{FM}$ is improved substantially with the embedding sizes increasing. Meanwhile, CL4CTR can improve the performance of FM with all embedding sizes. Furthermore, compared with the embedding size of 64 in FM, $CL4CTR_{FM}$ achieves better performance with a small size of 16. This means we can reduce the parameters while achieving better results by applying CL4CTR on FM.

\section{Conclusion}
In this paper, we propose a novel framework named Contrastive Learning for CTR prediction (CL4CTR), which directly improves the quality of feature representations, especially for low frequency features. In CL4CTR, we introduce a contrastive module to improve the quality and generalizability of the feature representations by fully utilizing the self-supervised signals from the features. Furthermore, considering the unique characteristics of CTR prediction tasks, we propose two constraints in contrastive learning: feature alignment and feature uniformity, which are used to regularize feature representations. The extensive experimental results demonstrate the excellent effectiveness and compatibility of our proposed CL4CTR on four public datasets.

\begin{acks}
This work was supported by the National Natural Science Foundation of China (NSFC) under Grants 62172106 and 61932007.
\end{acks}

\clearpage

\bibliographystyle{ACM-Reference-Format}
% \bibliography{CL4CTR.bbl}
% \bibliography{clfm_ref}

\begin{thebibliography}{46}

%%% the Web address.

\ifx \showCODEN    \undefined \def \showCODEN     #1{\unskip}     \fi
\ifx \showDOI      \undefined \def \showDOI       #1{#1}\fi
\ifx \showISBNx    \undefined \def \showISBNx     #1{\unskip}     \fi
\ifx \showISBNxiii \undefined \def \showISBNxiii  #1{\unskip}     \fi
\ifx \showISSN     \undefined \def \showISSN      #1{\unskip}     \fi
\ifx \showLCCN     \undefined \def \showLCCN      #1{\unskip}     \fi
\ifx \shownote     \undefined \def \shownote      #1{#1}          \fi
\ifx \showarticletitle \undefined \def \showarticletitle #1{#1}   \fi
\ifx \showURL      \undefined \def \showURL       {\relax}        \fi
% invisible to TeX
\providecommand\bibfield[2]{#2}
\providecommand\bibinfo[2]{#2}
\providecommand\natexlab[1]{#1}
\providecommand\showeprint[2][]{arXiv:#2}

\bibitem[Chen et~al\mbox{.}(2021a)]%
        {chen2021enhancing}
\bibfield{author}{\bibinfo{person}{Bo Chen}, \bibinfo{person}{Yichao Wang},
  \bibinfo{person}{Zhirong Liu}, \bibinfo{person}{Ruiming Tang},
  \bibinfo{person}{Wei Guo}, \bibinfo{person}{Hongkun Zheng},
  \bibinfo{person}{Weiwei Yao}, \bibinfo{person}{Muyu Zhang}, {and}
  \bibinfo{person}{Xiuqiang He}.} \bibinfo{year}{2021}\natexlab{a}.
\newblock \showarticletitle{Enhancing explicit and implicit feature
  interactions via information sharing for parallel deep CTR models}. In
  \bibinfo{booktitle}{\emph{CIKM}}. \bibinfo{pages}{3757--3766}.
\newblock


\bibitem[Chen et~al\mbox{.}(2020)]%
        {chen2020simple}
\bibfield{author}{\bibinfo{person}{Ting Chen}, \bibinfo{person}{Simon
  Kornblith}, \bibinfo{person}{Mohammad Norouzi}, {and}
  \bibinfo{person}{Geoffrey Hinton}.} \bibinfo{year}{2020}\natexlab{}.
\newblock \showarticletitle{A simple framework for contrastive learning of
  visual representations}. In \bibinfo{booktitle}{\emph{International
  conference on machine learning}}. PMLR, \bibinfo{pages}{1597--1607}.
\newblock


\bibitem[Chen et~al\mbox{.}(2021b)]%
        {chen2021dcap}
\bibfield{author}{\bibinfo{person}{Zekai Chen}, \bibinfo{person}{Fangtian
  Zhong}, \bibinfo{person}{Zhumin Chen}, \bibinfo{person}{Xiao Zhang},
  \bibinfo{person}{Robert Pless}, {and} \bibinfo{person}{Xiuzhen Cheng}.}
  \bibinfo{year}{2021}\natexlab{b}.
\newblock \showarticletitle{DCAP: Deep Cross Attentional Product Network for
  User Response Prediction}. In \bibinfo{booktitle}{\emph{Proceedings of the
  30th ACM International Conference on Information \& Knowledge Management}}.
  \bibinfo{pages}{221--230}.
\newblock


\bibitem[Cheng et~al\mbox{.}(2016)]%
        {cheng2016wide}
\bibfield{author}{\bibinfo{person}{Heng-Tze Cheng}, \bibinfo{person}{Levent
  Koc}, \bibinfo{person}{Jeremiah Harmsen}, \bibinfo{person}{Tal Shaked},
  \bibinfo{person}{Tushar Chandra}, \bibinfo{person}{Hrishi Aradhye},
  \bibinfo{person}{Glen Anderson}, \bibinfo{person}{Greg Corrado},
  \bibinfo{person}{Wei Chai}, \bibinfo{person}{Mustafa Ispir}, {et~al\mbox{.}}}
  \bibinfo{year}{2016}\natexlab{}.
\newblock \showarticletitle{Wide \& deep learning for recommender systems}. In
  \bibinfo{booktitle}{\emph{Proceedings of the 1st workshop on deep learning
  for recommender systems}}. \bibinfo{pages}{7--10}.
\newblock


\bibitem[Cheng et~al\mbox{.}(2020)]%
        {cheng2020adaptive}
\bibfield{author}{\bibinfo{person}{Weiyu Cheng}, \bibinfo{person}{Yanyan Shen},
  {and} \bibinfo{person}{Linpeng Huang}.} \bibinfo{year}{2020}\natexlab{}.
\newblock \showarticletitle{Adaptive factorization network: Learning
  adaptive-order feature interactions}. In
  \bibinfo{booktitle}{\emph{Proceedings of the AAAI Conference on Artificial
  Intelligence}}, Vol.~\bibinfo{volume}{34}. \bibinfo{pages}{3609--3616}.
\newblock


\bibitem[Gao et~al\mbox{.}(2021)]%
        {gao2021simcse}
\bibfield{author}{\bibinfo{person}{Tianyu Gao}, \bibinfo{person}{Xingcheng
  Yao}, {and} \bibinfo{person}{Danqi Chen}.} \bibinfo{year}{2021}\natexlab{}.
\newblock \showarticletitle{SimCSE: Simple Contrastive Learning of Sentence
  Embeddings}. In \bibinfo{booktitle}{\emph{Proceedings of the 2021 Conference
  on Empirical Methods in Natural Language Processing}}.
  \bibinfo{pages}{6894--6910}.
\newblock


\bibitem[Guo et~al\mbox{.}(2017)]%
        {guo2017deepfm}
\bibfield{author}{\bibinfo{person}{Huifeng Guo}, \bibinfo{person}{Ruiming
  Tang}, \bibinfo{person}{Yunming Ye}, \bibinfo{person}{Zhenguo Li}, {and}
  \bibinfo{person}{Xiuqiang He}.} \bibinfo{year}{2017}\natexlab{}.
\newblock \showarticletitle{DeepFM: a factorization-machine based neural
  network for CTR prediction}. In \bibinfo{booktitle}{\emph{Proceedings of the
  26th International Joint Conference on Artificial Intelligence}}.
\newblock


\bibitem[Guo et~al\mbox{.}(2022)]%
        {guo2022miss}
\bibfield{author}{\bibinfo{person}{Wei Guo}, \bibinfo{person}{Can Zhang},
  \bibinfo{person}{Zhicheng He}, \bibinfo{person}{Jiarui Qin},
  \bibinfo{person}{Huifeng Guo}, \bibinfo{person}{Bo Chen},
  \bibinfo{person}{Ruiming Tang}, \bibinfo{person}{Xiuqiang He}, {and}
  \bibinfo{person}{Rui Zhang}.} \bibinfo{year}{2022}\natexlab{}.
\newblock \showarticletitle{Miss: Multi-interest self-supervised learning
  framework for click-through rate prediction}. In
  \bibinfo{booktitle}{\emph{2022 IEEE 38th International Conference on Data
  Engineering (ICDE)}}. IEEE, \bibinfo{pages}{727--740}.
\newblock


\bibitem[He et~al\mbox{.}(2020)]%
        {he2020momentum}
\bibfield{author}{\bibinfo{person}{Kaiming He}, \bibinfo{person}{Haoqi Fan},
  \bibinfo{person}{Yuxin Wu}, \bibinfo{person}{Saining Xie}, {and}
  \bibinfo{person}{Ross Girshick}.} \bibinfo{year}{2020}\natexlab{}.
\newblock \showarticletitle{Momentum contrast for unsupervised visual
  representation learning}. In \bibinfo{booktitle}{\emph{Proceedings of the
  IEEE/CVF conference on computer vision and pattern recognition}}.
  \bibinfo{pages}{9729--9738}.
\newblock


\bibitem[He and Chua(2017)]%
        {he2017neural}
\bibfield{author}{\bibinfo{person}{Xiangnan He} {and} \bibinfo{person}{Tat-Seng
  Chua}.} \bibinfo{year}{2017}\natexlab{}.
\newblock \showarticletitle{Neural factorization machines for sparse predictive
  analytics}. In \bibinfo{booktitle}{\emph{Proceedings of the 40th
  International ACM SIGIR conference on Research and Development in Information
  Retrieval}}. \bibinfo{pages}{355--364}.
\newblock


\bibitem[Hinton et~al\mbox{.}(2012)]%
        {hinton2012improving}
\bibfield{author}{\bibinfo{person}{Geoffrey~E Hinton}, \bibinfo{person}{Nitish
  Srivastava}, \bibinfo{person}{Alex Krizhevsky}, \bibinfo{person}{Ilya
  Sutskever}, {and} \bibinfo{person}{Ruslan~R Salakhutdinov}.}
  \bibinfo{year}{2012}\natexlab{}.
\newblock \showarticletitle{Improving neural networks by preventing
  co-adaptation of feature detectors}.
\newblock \bibinfo{journal}{\emph{arXiv preprint arXiv:1207.0580}}
  (\bibinfo{year}{2012}).
\newblock


\bibitem[Huang et~al\mbox{.}(2020)]%
        {huang2020gatenet}
\bibfield{author}{\bibinfo{person}{Tongwen Huang}, \bibinfo{person}{Qingyun
  She}, \bibinfo{person}{Zhiqiang Wang}, {and} \bibinfo{person}{Junlin Zhang}.}
  \bibinfo{year}{2020}\natexlab{}.
\newblock \showarticletitle{GateNet: Gating-Enhanced Deep Network for
  Click-Through Rate Prediction}.
\newblock \bibinfo{journal}{\emph{arXiv preprint arXiv:2007.03519}}
  (\bibinfo{year}{2020}).
\newblock


\bibitem[Huang et~al\mbox{.}(2019)]%
        {huang2019fibinet}
\bibfield{author}{\bibinfo{person}{Tongwen Huang}, \bibinfo{person}{Zhiqi
  Zhang}, {and} \bibinfo{person}{Junlin Zhang}.}
  \bibinfo{year}{2019}\natexlab{}.
\newblock \showarticletitle{FiBiNET: combining feature importance and bilinear
  feature interaction for click-through rate prediction}. In
  \bibinfo{booktitle}{\emph{Proceedings of the 13th ACM Conference on
  Recommender Systems}}. \bibinfo{pages}{169--177}.
\newblock


\bibitem[Juan et~al\mbox{.}(2016)]%
        {juan2016field}
\bibfield{author}{\bibinfo{person}{Yuchin Juan}, \bibinfo{person}{Yong Zhuang},
  \bibinfo{person}{Wei-Sheng Chin}, {and} \bibinfo{person}{Chih-Jen Lin}.}
  \bibinfo{year}{2016}\natexlab{}.
\newblock \showarticletitle{Field-aware factorization machines for CTR
  prediction}. In \bibinfo{booktitle}{\emph{Proceedings of the 10th ACM
  Conference on Recommender Systems}}. \bibinfo{pages}{43--50}.
\newblock


\bibitem[Lee et~al\mbox{.}(2021)]%
        {lee2021bootstrapping}
\bibfield{author}{\bibinfo{person}{Dongha Lee}, \bibinfo{person}{SeongKu Kang},
  \bibinfo{person}{Hyunjun Ju}, \bibinfo{person}{Chanyoung Park}, {and}
  \bibinfo{person}{Hwanjo Yu}.} \bibinfo{year}{2021}\natexlab{}.
\newblock \showarticletitle{Bootstrapping user and item representations for
  one-class collaborative filtering}. In \bibinfo{booktitle}{\emph{Proceedings
  of the 44th International ACM SIGIR Conference on Research and Development in
  Information Retrieval}}. \bibinfo{pages}{317--326}.
\newblock


\bibitem[Li et~al\mbox{.}(2020)]%
        {li2020interpretable}
\bibfield{author}{\bibinfo{person}{Zeyu Li}, \bibinfo{person}{Wei Cheng},
  \bibinfo{person}{Yang Chen}, \bibinfo{person}{Haifeng Chen}, {and}
  \bibinfo{person}{Wei Wang}.} \bibinfo{year}{2020}\natexlab{}.
\newblock \showarticletitle{Interpretable click-through rate prediction through
  hierarchical attention}. In \bibinfo{booktitle}{\emph{Proceedings of the 13th
  International Conference on Web Search and Data Mining}}.
  \bibinfo{pages}{313--321}.
\newblock


\bibitem[Lian et~al\mbox{.}(2018)]%
        {lian2018xdeepfm}
\bibfield{author}{\bibinfo{person}{Jianxun Lian}, \bibinfo{person}{Xiaohuan
  Zhou}, \bibinfo{person}{Fuzheng Zhang}, \bibinfo{person}{Zhongxia Chen},
  \bibinfo{person}{Xing Xie}, {and} \bibinfo{person}{Guangzhong Sun}.}
  \bibinfo{year}{2018}\natexlab{}.
\newblock \showarticletitle{xdeepfm: Combining explicit and implicit feature
  interactions for recommender systems}. In
  \bibinfo{booktitle}{\emph{Proceedings of the 24th ACM SIGKDD International
  Conference on Knowledge Discovery \& Data Mining}}.
  \bibinfo{pages}{1754--1763}.
\newblock


\bibitem[Liu et~al\mbox{.}(2019)]%
        {liu2019feature}
\bibfield{author}{\bibinfo{person}{Bin Liu}, \bibinfo{person}{Ruiming Tang},
  \bibinfo{person}{Yingzhi Chen}, \bibinfo{person}{Jinkai Yu},
  \bibinfo{person}{Huifeng Guo}, {and} \bibinfo{person}{Yuzhou Zhang}.}
  \bibinfo{year}{2019}\natexlab{}.
\newblock \showarticletitle{Feature generation by convolutional neural network
  for click-through rate prediction}. In \bibinfo{booktitle}{\emph{The World
  Wide Web Conference}}. \bibinfo{pages}{1119--1129}.
\newblock


\bibitem[Lu et~al\mbox{.}(2020)]%
        {lu2020dual}
\bibfield{author}{\bibinfo{person}{Wantong Lu}, \bibinfo{person}{Yantao Yu},
  \bibinfo{person}{Yongzhe Chang}, \bibinfo{person}{Zhen Wang},
  \bibinfo{person}{Chenhui Li}, {and} \bibinfo{person}{Bo Yuan}.}
  \bibinfo{year}{2020}\natexlab{}.
\newblock \showarticletitle{A Dual Input-aware Factorization Machine for CTR
  Prediction}. In \bibinfo{booktitle}{\emph{IJCAI}}.
  \bibinfo{pages}{3139--3145}.
\newblock


\bibitem[Luo et~al\mbox{.}(2020)]%
        {luo2020network}
\bibfield{author}{\bibinfo{person}{Yuanfei Luo}, \bibinfo{person}{Hao Zhou},
  \bibinfo{person}{Wei-Wei Tu}, \bibinfo{person}{Yuqiang Chen},
  \bibinfo{person}{Wenyuan Dai}, {and} \bibinfo{person}{Qiang Yang}.}
  \bibinfo{year}{2020}\natexlab{}.
\newblock \showarticletitle{Network on network for tabular data classification
  in real-world applications}. In \bibinfo{booktitle}{\emph{Proceedings of the
  43rd International ACM SIGIR Conference on Research and Development in
  Information Retrieval}}. \bibinfo{pages}{2317--2326}.
\newblock


\bibitem[Meng et~al\mbox{.}(2021)]%
        {meng2021general}
\bibfield{author}{\bibinfo{person}{Ze Meng}, \bibinfo{person}{Jinnian Zhang},
  \bibinfo{person}{Yumeng Li}, \bibinfo{person}{Jiancheng Li},
  \bibinfo{person}{Tanchao Zhu}, {and} \bibinfo{person}{Lifeng Sun}.}
  \bibinfo{year}{2021}\natexlab{}.
\newblock \showarticletitle{A general method for automatic discovery of
  powerful interactions in click-through rate prediction}. In
  \bibinfo{booktitle}{\emph{Proceedings of the 44th International ACM SIGIR
  Conference on Research and Development in Information Retrieval}}.
  \bibinfo{pages}{1298--1307}.
\newblock


\bibitem[Pan et~al\mbox{.}(2018)]%
        {pan2018field}
\bibfield{author}{\bibinfo{person}{Junwei Pan}, \bibinfo{person}{Jian Xu},
  \bibinfo{person}{Alfonso~Lobos Ruiz}, \bibinfo{person}{Wenliang Zhao},
  \bibinfo{person}{Shengjun Pan}, \bibinfo{person}{Yu Sun}, {and}
  \bibinfo{person}{Quan Lu}.} \bibinfo{year}{2018}\natexlab{}.
\newblock \showarticletitle{Field-weighted factorization machines for
  click-through rate prediction in display advertising}. In
  \bibinfo{booktitle}{\emph{Proceedings of the 2018 World Wide Web
  Conference}}. \bibinfo{pages}{1349--1357}.
\newblock


\bibitem[Pan et~al\mbox{.}(2021)]%
        {pan2021aqclclick}
\bibfield{author}{\bibinfo{person}{Yujie Pan}, \bibinfo{person}{Jiangchao Yao},
  \bibinfo{person}{Bo Han}, \bibinfo{person}{Kunyang Jia}, \bibinfo{person}{Ya
  Zhang}, {and} \bibinfo{person}{Hongxia Yang}.}
  \bibinfo{year}{2021}\natexlab{}.
\newblock \showarticletitle{Click-through Rate Prediction with Auto-Quantized
  Contrastive Learning}.
\newblock \bibinfo{journal}{\emph{arXiv preprint arXiv:2109.13921}}
  (\bibinfo{year}{2021}).
\newblock


\bibitem[Qin et~al\mbox{.}(2020)]%
        {qin2020user}
\bibfield{author}{\bibinfo{person}{Jiarui Qin}, \bibinfo{person}{Weinan Zhang},
  \bibinfo{person}{Xin Wu}, \bibinfo{person}{Jiarui Jin},
  \bibinfo{person}{Yuchen Fang}, {and} \bibinfo{person}{Yong Yu}.}
  \bibinfo{year}{2020}\natexlab{}.
\newblock \showarticletitle{User behavior retrieval for click-through rate
  prediction}. In \bibinfo{booktitle}{\emph{Proceedings of the 43rd
  International ACM SIGIR Conference on Research and Development in Information
  Retrieval}}. \bibinfo{pages}{2347--2356}.
\newblock


\bibitem[Qu et~al\mbox{.}(2018)]%
        {qu2018product}
\bibfield{author}{\bibinfo{person}{Yanru Qu}, \bibinfo{person}{Bohui Fang},
  \bibinfo{person}{Weinan Zhang}, \bibinfo{person}{Ruiming Tang},
  \bibinfo{person}{Minzhe Niu}, \bibinfo{person}{Huifeng Guo},
  \bibinfo{person}{Yong Yu}, {and} \bibinfo{person}{Xiuqiang He}.}
  \bibinfo{year}{2018}\natexlab{}.
\newblock \showarticletitle{Product-based neural networks for user response
  prediction over multi-field categorical data}.
\newblock \bibinfo{journal}{\emph{ACM Transactions on Information Systems
  (TOIS)}} \bibinfo{volume}{37}, \bibinfo{number}{1} (\bibinfo{year}{2018}),
  \bibinfo{pages}{1--35}.
\newblock


\bibitem[Rendle(2012)]%
        {rendle2012factorization}
\bibfield{author}{\bibinfo{person}{Steffen Rendle}.}
  \bibinfo{year}{2012}\natexlab{}.
\newblock \showarticletitle{Factorization machines with libfm}.
\newblock \bibinfo{journal}{\emph{ACM Transactions on Intelligent Systems and
  Technology (TIST)}} \bibinfo{volume}{3}, \bibinfo{number}{3}
  (\bibinfo{year}{2012}), \bibinfo{pages}{1--22}.
\newblock


\bibitem[Richardson et~al\mbox{.}(2007)]%
        {richardson2007predicting}
\bibfield{author}{\bibinfo{person}{Matthew Richardson}, \bibinfo{person}{Ewa
  Dominowska}, {and} \bibinfo{person}{Robert Ragno}.}
  \bibinfo{year}{2007}\natexlab{}.
\newblock \showarticletitle{Predicting clicks: estimating the click-through
  rate for new ads}. In \bibinfo{booktitle}{\emph{Proceedings of the 16th
  international conference on World Wide Web}}. \bibinfo{pages}{521--530}.
\newblock


\bibitem[Song et~al\mbox{.}(2019)]%
        {song2019autoint}
\bibfield{author}{\bibinfo{person}{Weiping Song}, \bibinfo{person}{Chence Shi},
  \bibinfo{person}{Zhiping Xiao}, \bibinfo{person}{Zhijian Duan},
  \bibinfo{person}{Yewen Xu}, \bibinfo{person}{Ming Zhang}, {and}
  \bibinfo{person}{Jian Tang}.} \bibinfo{year}{2019}\natexlab{}.
\newblock \showarticletitle{Autoint: Automatic feature interaction learning via
  self-attentive neural networks}. In \bibinfo{booktitle}{\emph{Proceedings of
  the 28th ACM International Conference on Information and Knowledge
  Management}}. \bibinfo{pages}{1161--1170}.
\newblock


\bibitem[Sun et~al\mbox{.}(2021)]%
        {sun2021fm2}
\bibfield{author}{\bibinfo{person}{Yang Sun}, \bibinfo{person}{Junwei Pan},
  \bibinfo{person}{Alex Zhang}, {and} \bibinfo{person}{Aaron Flores}.}
  \bibinfo{year}{2021}\natexlab{}.
\newblock \showarticletitle{Fm2: Field-matrixed factorization machines for
  recommender systems}. In \bibinfo{booktitle}{\emph{Proceedings of the Web
  Conference 2021}}. \bibinfo{pages}{2828--2837}.
\newblock


\bibitem[Vaswani et~al\mbox{.}(2017)]%
        {vaswani2017attention}
\bibfield{author}{\bibinfo{person}{Ashish Vaswani}, \bibinfo{person}{Noam
  Shazeer}, \bibinfo{person}{Niki Parmar}, \bibinfo{person}{Jakob Uszkoreit},
  \bibinfo{person}{Llion Jones}, \bibinfo{person}{Aidan~N Gomez},
  \bibinfo{person}{{\L}ukasz Kaiser}, {and} \bibinfo{person}{Illia
  Polosukhin}.} \bibinfo{year}{2017}\natexlab{}.
\newblock \showarticletitle{Attention is all you need}. In
  \bibinfo{booktitle}{\emph{Advances in neural information processing
  systems}}. \bibinfo{pages}{5998--6008}.
\newblock


\bibitem[Verma et~al\mbox{.}(2021)]%
        {verma2021towards}
\bibfield{author}{\bibinfo{person}{Vikas Verma}, \bibinfo{person}{Thang Luong},
  \bibinfo{person}{Kenji Kawaguchi}, \bibinfo{person}{Hieu Pham}, {and}
  \bibinfo{person}{Quoc Le}.} \bibinfo{year}{2021}\natexlab{}.
\newblock \showarticletitle{Towards domain-agnostic contrastive learning}. In
  \bibinfo{booktitle}{\emph{International Conference on Machine Learning}}.
  PMLR, \bibinfo{pages}{10530--10541}.
\newblock


\bibitem[Wang and Liu(2021)]%
        {wang2021understanding2}
\bibfield{author}{\bibinfo{person}{Feng Wang} {and} \bibinfo{person}{Huaping
  Liu}.} \bibinfo{year}{2021}\natexlab{}.
\newblock \showarticletitle{Understanding the behaviour of contrastive loss}.
  In \bibinfo{booktitle}{\emph{Proceedings of the IEEE/CVF conference on
  computer vision and pattern recognition}}. \bibinfo{pages}{2495--2504}.
\newblock


\bibitem[Wang et~al\mbox{.}(2022)]%
        {wang2022enhancing}
\bibfield{author}{\bibinfo{person}{Fangye Wang}, \bibinfo{person}{Yingxu Wang},
  \bibinfo{person}{Dongsheng Li}, \bibinfo{person}{Hansu Gu},
  \bibinfo{person}{Tun Lu}, \bibinfo{person}{Peng Zhang}, {and}
  \bibinfo{person}{Ning Gu}.} \bibinfo{year}{2022}\natexlab{}.
\newblock \showarticletitle{Enhancing CTR Prediction with Context-Aware Feature
  Representation Learning}. In \bibinfo{booktitle}{\emph{Proceedings of the
  44th International ACM SIGIR Conference on Research and Development in
  Information Retrieval}}.
\newblock


\bibitem[Wang et~al\mbox{.}(2017)]%
        {wang2017deep}
\bibfield{author}{\bibinfo{person}{Ruoxi Wang}, \bibinfo{person}{Bin Fu},
  \bibinfo{person}{Gang Fu}, {and} \bibinfo{person}{Mingliang Wang}.}
  \bibinfo{year}{2017}\natexlab{}.
\newblock \showarticletitle{Deep \& cross network for ad click predictions}.
\newblock In \bibinfo{booktitle}{\emph{Proceedings of the ADKDD'17}}.
  \bibinfo{pages}{1--7}.
\newblock


\bibitem[Wang et~al\mbox{.}(2020)]%
        {wang2020dcn}
\bibfield{author}{\bibinfo{person}{Ruoxi Wang}, \bibinfo{person}{Rakesh
  Shivanna}, \bibinfo{person}{Derek~Z Cheng}, \bibinfo{person}{Sagar Jain},
  \bibinfo{person}{Dong Lin}, \bibinfo{person}{Lichan Hong}, {and}
  \bibinfo{person}{Ed~H Chi}.} \bibinfo{year}{2020}\natexlab{}.
\newblock \showarticletitle{DCN-M: Improved Deep \& Cross Network for Feature
  Cross Learning in Web-scale Learning to Rank Systems}.
\newblock \bibinfo{journal}{\emph{arXiv preprint arXiv:2008.13535}}
  (\bibinfo{year}{2020}).
\newblock


\bibitem[Wang and Isola(2020)]%
        {wang2020understanding}
\bibfield{author}{\bibinfo{person}{Tongzhou Wang} {and}
  \bibinfo{person}{Phillip Isola}.} \bibinfo{year}{2020}\natexlab{}.
\newblock \showarticletitle{Understanding contrastive representation learning
  through alignment and uniformity on the hypersphere}. In
  \bibinfo{booktitle}{\emph{International Conference on Machine Learning}}.
  PMLR, \bibinfo{pages}{9929--9939}.
\newblock


\bibitem[Wang et~al\mbox{.}(2021)]%
        {wang2021masknet}
\bibfield{author}{\bibinfo{person}{Zhiqiang Wang}, \bibinfo{person}{Qingyun
  She}, {and} \bibinfo{person}{Junlin Zhang}.} \bibinfo{year}{2021}\natexlab{}.
\newblock \showarticletitle{MaskNet: introducing feature-wise multiplication to
  CTR ranking models by instance-guided mask}.
\newblock \bibinfo{journal}{\emph{arXiv preprint arXiv:2102.07619}}
  (\bibinfo{year}{2021}).
\newblock


\bibitem[Wu et~al\mbox{.}(2020)]%
        {wu2020tfnet}
\bibfield{author}{\bibinfo{person}{Shu Wu}, \bibinfo{person}{Feng Yu},
  \bibinfo{person}{Xueli Yu}, \bibinfo{person}{Qiang Liu},
  \bibinfo{person}{Liang Wang}, \bibinfo{person}{Tieniu Tan},
  \bibinfo{person}{Jie Shao}, {and} \bibinfo{person}{Fan Huang}.}
  \bibinfo{year}{2020}\natexlab{}.
\newblock \showarticletitle{TFNet: Multi-Semantic Feature Interaction for CTR
  Prediction}. In \bibinfo{booktitle}{\emph{Proceedings of the 43rd
  International ACM SIGIR Conference on Research and Development in Information
  Retrieval}}. \bibinfo{pages}{1885--1888}.
\newblock


\bibitem[Xiao et~al\mbox{.}(2017)]%
        {xiao2017attentional}
\bibfield{author}{\bibinfo{person}{Jun Xiao}, \bibinfo{person}{Hao Ye},
  \bibinfo{person}{Xiangnan He}, \bibinfo{person}{Hanwang Zhang},
  \bibinfo{person}{Fei Wu}, {and} \bibinfo{person}{Tat-Seng Chua}.}
  \bibinfo{year}{2017}\natexlab{}.
\newblock \showarticletitle{Attentional Factorization Machines: Learning the
  Weight of Feature Interactions via Attention Networks}. In
  \bibinfo{booktitle}{\emph{IJCAI}}.
\newblock


\bibitem[Xie et~al\mbox{.}(2022)]%
        {xie2020CL4SREC}
\bibfield{author}{\bibinfo{person}{Xu Xie}, \bibinfo{person}{Fei Sun},
  \bibinfo{person}{Zhaoyang Liu}, \bibinfo{person}{Shiwen Wu},
  \bibinfo{person}{Jinyang Gao}, \bibinfo{person}{Jiandong Zhang},
  \bibinfo{person}{Bolin Ding}, {and} \bibinfo{person}{Bin Cui}.}
  \bibinfo{year}{2022}\natexlab{}.
\newblock \showarticletitle{Contrastive learning for sequential
  recommendation}. In \bibinfo{booktitle}{\emph{2022 IEEE 38th International
  Conference on Data Engineering (ICDE)}}. IEEE.
\newblock


\bibitem[Yu et~al\mbox{.}(2022)]%
        {yu2022self}
\bibfield{author}{\bibinfo{person}{Junliang Yu}, \bibinfo{person}{Hongzhi Yin},
  \bibinfo{person}{Xin Xia}, \bibinfo{person}{Tong Chen},
  \bibinfo{person}{Jundong Li}, {and} \bibinfo{person}{Zi Huang}.}
  \bibinfo{year}{2022}\natexlab{}.
\newblock \showarticletitle{Self-Supervised Learning for Recommender Systems: A
  Survey}.
\newblock \bibinfo{journal}{\emph{arXiv preprint arXiv:2203.15876}}
  (\bibinfo{year}{2022}).
\newblock


\bibitem[Yu et~al\mbox{.}(2019)]%
        {yu2019input}
\bibfield{author}{\bibinfo{person}{Yantao Yu}, \bibinfo{person}{Zhen Wang},
  {and} \bibinfo{person}{Bo Yuan}.} \bibinfo{year}{2019}\natexlab{}.
\newblock \showarticletitle{An Input-aware Factorization Machine for Sparse
  Prediction}. In \bibinfo{booktitle}{\emph{IJCAI}}.
  \bibinfo{pages}{1466--1472}.
\newblock


\bibitem[Zhang et~al\mbox{.}(2021)]%
        {zhang2021survey_deep}
\bibfield{author}{\bibinfo{person}{Weinan Zhang}, \bibinfo{person}{Jiarui Qin},
  \bibinfo{person}{Wei Guo}, \bibinfo{person}{Ruiming Tang}, {and}
  \bibinfo{person}{Xiuqiang He}.} \bibinfo{year}{2021}\natexlab{}.
\newblock \showarticletitle{Deep learning for click-through rate estimation}.
  In \bibinfo{booktitle}{\emph{IJCAI}}.
\newblock


\bibitem[Zhao et~al\mbox{.}(2020)]%
        {zhao2020fed_dimension}
\bibfield{author}{\bibinfo{person}{Zihao Zhao}, \bibinfo{person}{Zhiwei Fang},
  \bibinfo{person}{Yong Li}, \bibinfo{person}{Changping Peng},
  \bibinfo{person}{Yongjun Bao}, {and} \bibinfo{person}{Weipeng Yan}.}
  \bibinfo{year}{2020}\natexlab{}.
\newblock \showarticletitle{Dimension Relation Modeling for Click-Through Rate
  Prediction}. In \bibinfo{booktitle}{\emph{Proceedings of the 29th ACM
  International Conference on Information \& Knowledge Management}}.
  \bibinfo{pages}{2333--2336}.
\newblock


\bibitem[Zhao et~al\mbox{.}(2022)]%
        {zhao2022fint}
\bibfield{author}{\bibinfo{person}{Zhishan Zhao}, \bibinfo{person}{Sen Yang},
  \bibinfo{person}{Guohui Liu}, \bibinfo{person}{Dawei Feng}, {and}
  \bibinfo{person}{Kele Xu}.} \bibinfo{year}{2022}\natexlab{}.
\newblock \showarticletitle{FINT: Field-Aware Interaction Neural Network for
  Click-Through Rate Prediction}. In \bibinfo{booktitle}{\emph{ICASSP 2022-2022
  IEEE International Conference on Acoustics, Speech and Signal Processing
  (ICASSP)}}. IEEE, \bibinfo{pages}{3913--3917}.
\newblock


\bibitem[Zhou et~al\mbox{.}(2021)]%
        {zhou2021selfcf}
\bibfield{author}{\bibinfo{person}{Xin Zhou}, \bibinfo{person}{Aixin Sun},
  \bibinfo{person}{Yong Liu}, \bibinfo{person}{Jie Zhang}, {and}
  \bibinfo{person}{Chunyan Miao}.} \bibinfo{year}{2021}\natexlab{}.
\newblock \showarticletitle{SelfCF: A Simple Framework for Self-supervised
  Collaborative Filtering}.
\newblock \bibinfo{journal}{\emph{arXiv preprint arXiv:2107.03019}}
  (\bibinfo{year}{2021}).
\newblock


\end{thebibliography}
\end{document}